%% file: cpsh08.tex
\documentclass[12pt,a4paper]{article}
\usepackage{color,graphics,amsmath,epsfig,rotating,cite}

\begin{document}

\newcommand\beqn{\begin{eqnarray}}
\newcommand\eeqn{\end{eqnarray}}
\newcommand{\ra}{\rightarrow}

\def\tb{\tan\beta}

\def\x{\chi}
\def\ti{\tilde}
\def\nt{\tilde \x^0}
\def\ch{\tilde \x^+}
\def\chpm{\tilde \x^\pm}
\def\st{\tilde t}
\def\stau{\tilde\tau}
\def\noi{\noindent}

\def\sigv{\langle\sigma v\rangle}

\def\micromegas{{\tt micrOMEGAs\,2.0}}
\def\micro{{\tt micrOMEGAs}}
\def\darksusy{{\tt DarkSUSY}}
\def\isared{{\tt IsaRED}}

\newcommand{\mnt}[1]   {m_{\tilde\x^0_{#1}}}
\newcommand{\mch}[1]   {m_{\tilde\x^\pm_{#1}}}
\newcommand{\msf}[1]   {m_{\tilde f_{#1}}}
\newcommand{\mst}[1]   {m_{\tilde t_{#1}}}
\newcommand{\mstau}[1] {m_{\tilde\tau_{#1}}}

\newcommand{\eq}[1]  {\mbox{(\ref{eq:#1})}}
\newcommand{\fig}[1] {Fig.~\ref{fig:#1}}
\newcommand{\Fig}[1] {Figure~\ref{fig:#1}}
\newcommand{\tab}[1] {Table~\ref{tab:#1}}
\newcommand{\Tab}[1] {Table~\ref{tab:#1}}

\newcommand{\ablabels}[3]{
  \begin{picture}(100,0)\setlength{\unitlength}{1mm}
    \put(#1,#3){\bf (a)}
    \put(#2,#3){\bf (b)}
  \end{picture}\\[-8mm]
}

\newcommand{\gsim}{\;\raisebox{-0.9ex}
           {$\textstyle\stackrel{\textstyle >}{\sim}$}\;}
\newcommand{\lsim}{\;\raisebox{-0.9ex}{$\textstyle\stackrel{\textstyle<}
           {\sim}$}\;}


\vspace*{-18mm}
\begin{flushright}
  LAPTH-1145/06\\
  CERN-PH-TH/2006-063\\
  hep-ph/0604150\\
\end{flushright}
\vspace*{4mm}

\begin{center}

{\Large\bf
   Relic density of neutralino dark matter\\[3mm]
   in the MSSM with CP violation}\\[6mm]

{\large
   G. B\'elanger$^1$, F. Boudjema$^1$,
   S. Kraml$^2$, A. Pukhov$^3$, A. Semenov$^4$}\\[6mm]

{\it
1) LAPTH, 9 Chemin de Bellevue, B.P. 110, F-74941 Annecy-le-Vieux , France\\
2) CERN, Dept. of Physics, Theory Division, CH-1211 Geneva 23, Switzerland\\
3) Skobeltsyn Inst. of Nuclear Physics, Moscow State Univ., Moscow 119992, Russia\\
4) Joint Institute for Nuclear Research (JINR), 141980, Dubna, Russia}\\[6mm]

\end{center}

\begin{abstract}
We calculate the relic density of dark matter in the MSSM with CP
violation. We analyse various scenarios of neutralino
annihilation: the cases of a bino, bino-wino and bino-Higgsino
LSP, annihilation through Higgs, as well as sfermion
coannihilation scenarios. Large phase effects are found, on the
one hand due to shifts in the masses, on the other hand due to
modifications of the couplings. Taking special care to disentangle
the effects in masses and couplings, we demonstrate that the
presence of CP phases can have a significant influence on the
neutralino relic abundance. Typical variations in $\Omega h^2$
solely from modifications in the couplings are ${\cal O}$(10\%--100\%),
but can reach an order of magnitude in some cases.
\end{abstract}


\tableofcontents

\section{Introduction}

With the conclusive evidence for a significant component of cold dark
matter (CDM)
in the Universe, there is considerable interest, both at
the theoretical and experimental level, to identify this CDM and analyse
its properties; see \cite{Bertone:2004pz} for a recent review.
In particular, if the CDM consists of a new weakly interacting massive
particle (WIMP), as predicted in generic new physics models with
a discrete symmetry that ensures the stability of the lightest
particle, the next generation of colliders has good prospects to
discover it.
Being electrically neutral and stable, the WIMP escapes
the detector as missing energy and momentum.
The preferred discovery channels therefore rely on the production
of other new particles present in the theory and their decays into
the CDM candidate. By measuring the properties and decay
kinematics of these new particles, one should then be able to
determine the properties of the WIMP. If the measurements are
precise enough, this allows to predict the annihilation cross
sections and hence the thermal relic density of the CDM
candidate, thus checking the consistency between a particular
model of new physics and cosmology.

Indeed,  particle physics models trying to explain the dark matter
are constrained by recent precision cosmological measurements.
These are most notably the data from WMAP
\cite{Bennett:2003bz,Spergel:2003cb} and SDSS
\cite{Tegmark:2003ud}, which imply a (dominantly) cold dark matter
density of
\begin{equation}
  0.0945 < \Omega_{\rm CDM} h^2 < 0.1287
  \label{eq:wmap}
\end{equation}
at $2\sigma$.
In the following we will refer to Eq.~\eq{wmap} as the WMAP range.

The relic density of dark matter has been discussed extensively in
the framework of the most popular model for new physics, low-scale
supersymmetry (SUSY) with R-parity conservation. Especially if the
lightest supersymmetric particle (LSP) is the lightest neutralino,
this provides a good 
cold dark matter candidate
\cite{Goldberg:1983nd,Ellis:1983ew}. For the standard picture of
thermal freeze-out \cite{Kolb:1990vq}, 
assuming no additional non-thermal production mechanism,
comprehensive public codes that compute the relic density 
of the neutralino LSP are available today:
\micro~\cite{Belanger:2004yn,Belanger:2001fz},
\darksusy~\cite{Gondolo:2004sc}, and \isared~\cite{Baer:2002fv}.
Many analyses of neutralino dark matter were performed in the
minimal supersymmetric standard model (MSSM), whose parameters are
defined at the weak scale, see
e.g.~\cite{Profumo:2004at,Allanach:2004xn,Arkani-Hamed:2006mb}, in
the constrained MSSM (CMSSM) and mSUGRA models
~\cite{Drees:1992am,Roszkowski:2001sb,Ellis:2003cw,Pallis:2003jc,Baer:2003yh,
Lahanas:2003yz,Chattopadhyay:2003xi,Baltz:2004aw,Belanger:2005jk,
Allanach:2005kz,Djouadi:2006be}, 
and in other models with the parameters defined
at the unification scale \cite{Arnowitt:2001yh,Ellis:2002iu,
Birkedal-Hansen:2002am,Bertin:2002sq,Binetruy:2003yf,
Baer:2004fu,Belanger:2004ag,Baer:2005bu,King:2006tf}.

These studies showed that in the MSSM there are only a few
mechanisms that provide the correct amount of  neutralino
annihilation, consistent with the WMAP range Eq.~\eq{wmap}:
annihilation of a bino LSP into fermion pairs through $t$-channel
sfermion exchange in case of very light sparticles;
annihilation of a mixed bino-Higgsino or bino-wino LSP into gauge
boson pairs through $t$-channel chargino and neutralino exchange,
and into top-quark pairs through $s$-channel $Z$ exchange;
and finally annihilation near a Higgs resonance (the so-called Higgs funnel).
Furthermore, coannihilation processes with sparticles that are close
in mass with the LSP may bring $\Omega h^2$ in the desired range.
In particular, coannihilation with light sfermions can help to reduce
the relic density of a bino-like LSP.
Here note that coannihilation generically occurs when there is
a small mass gap between the LSP and the next-to-lightest SUSY particle (NLSP).
In scenarios with a Higgsino or wino LSP, one has in fact a mass-degenerate
triplet of Higgsinos or winos, and coannihilations are so efficient that
$\Omega h^2$ turns out much too small, unless the LSP has a mass of
order TeV.
In the case that $\Omega h^2$ is too low one would need, for instance,
a significant contribution from non-thermal production.

Most of these previous analyses assumed that CP is conserved,
although CP-violating phases are generic in the MSSM.
Here note that, given the Higgs mass bound $m_h > 114$~GeV
from LEP \cite{Barate:2003sz},
the CKM description of CP violation (CPV) in the SM is not
sufficient to generate the correct baryon asymmetry in the Universe.
One possible solution to this problem is leptogenesis. Another solution
is electroweak baryogenesis through new sources of CP violation beyond the SM.
Electroweak baryogenesis in the MSSM with CP violation (CPV-MSSM) has
been studied in \cite{Balazs:2004ae,Konstandin:2005cd,Cirigliano:2006dg};
see also references therein.
For a strongly first-order phase transition, it relies on the
existence of a light
stop~\cite{Delepine:1996vn,Balazs:2004bu,Cline:1998hy,Carena:1997ki}.

The parameters that can have CP phases are the gaugino
and Higgsino mass parameters and the trilinear sfermion-Higgs couplings.
Although constrained by electric dipole moments, nonzero phases can
significantly influence the phenomenology of SUSY particles.
They can also have a strong impact on the Higgs sector, inducing
scalar-pseudoscalar mixing through loop 
effects~\cite{Pilaftsis:1998dd,Pilaftsis:1999qt}.
This can in turn have a potentially dramatic effect on the 
relic density prediction in the Higgs funnel region:
neutralino annnihilation through $s$-channel scalar exchange is $p$-wave
suppressed; at small velocities it is dominated by pseudoscalar exchange.
In the presence of phases, both heavy Higgs bosons can, for instance,
acquire a pseudoscalar component and hence significantly contribute
to neutralino annihilation even at small velocities. Likewise, when
only one of the resonances is accessible to the neutralino annihilation,
large effects can be expected by changing the scalar/pseudoscalar content
of this resonance.
Last but not least, the couplings of the LSP to other sparticles depend
on the phases, and so will all the annihilation and coannihilation
cross sections, even though this is not a CP-violating (CP-odd) effect.

Consequences of complex parameters for the relic density of neutralinos
have so far been considered in the literature only for specific cases.
The effect of a phase in the squark left--right mixing was shown
to enhance the cross section for neutralino annihilation into
fermion pairs~\cite{Falk:1995fk}, but the important sfermion
coannihilation effects were not included.
In \cite{Nihei:2004bc}, the influence of phases was discussed
for various cases gaugino-Higgsino mixing, taking into account
CPV in the Higgs sector. However, coannihilations were again disregarded.
Reference~\cite{Argyrou:2004cs} discussed phase effects in neutralino
annihilation through Higgs exchange in the CMSSM at large $\tan\beta$.
Likewise, Ref.~\cite{Gomez:2005nr} examined a very specific case, namely
the dependence of $\Omega h^2$ on phases that arise through
supersymmetric loop corrections to the b quark mass
in SUGRA models with Yukawa unification.
For the general MSSM, the effect of a phase in the trilinear coupling
of the top squark on the neutralino annihilation near a Higgs resonance
were discussed in \cite{Gondolo:1999gu} and more recently in
\cite{Choi:2006hi}, however without analysing the phase dependence
of $\Omega h^2$.
Finally, the case of coannihilation with very light stops was discussed
in \cite{Balazs:2004ae,Cirigliano:2006dg} in the framework of electroweak
baryogenesis. Also these latter papers considered only few fixed values of
$\phi_\mu$ and did not study the phase dependence of $\Omega h^2$ in detail.
It is, moreover, important to note that all the above mentioned
publications studied the impact of phases solely in terms of
SUSY-breaking parameters, without discussing that part of the
phase effects are due to changes in the masses of the involved
particles.

A complete and coherent analysis of the relic density of
neutralinos in the CPV-MSSM is therefore still missing, and this
is the gap we intend to fill with this paper. We perform a
comprehensive analysis of the impact of CP phases on the relic
density of neutralino dark matter, taking into account
consistently phases in ALL annihilation and coannihilation
channels. To this end, we realized the implementation of the CPV-MSSM
within \micromegas~\cite{mo2:Allanach:2006fy,omcp:Allanach:2006fy}.
For the computation of masses, mixings and effective couplings in
the Higgs sector, we rely on {\tt CPsuperH}~\cite{Lee:2003nt}.
We further take into account collider constraints from sparticle and 
Higgs searches, as well as the constraint arising from the electric 
dipole moment of the electron~\cite{Hagiwara:2002fs} .

Besides realising the complete calculation of the relic density of
neutralinos in the CPV-MSSM, we analyse in detail the
above-mentioned scenarios of neutralino annihilation and
coanihilation, for which the LSP is a `good' CDM candidate. We
find indeed a large impact from phases due to modifications in the
sparticle couplings (through changes in the mixing matrices) but
also due to changes in the physical masses. Some of the
largest effects are in fact due to kinematics.
This should be expected as the relic density is often very
sensitive to masses, in particular 
to the exact mass difference between the LSP and NLSP in
coannihilation processes, or in the case of annihilation near a
$s$-channel Higgs resonance, to the difference between twice the
LSP mass and the mass of the Higgs. In these scenarios, setting
apart the purely kinematic effects hence somewhat tames the huge
effects due to CP phases found in some of the previous studies
listed above. On the other hand, we also find cases where the
phase dependences of masses and couplings work against each other,
and taking out the kinematic effects actually enhances the phase
dependence of the number density. There are even cases in which,
for fixed MSSM parameters, $\Omega h^2$ decreases with increasing
phase, but once the masses are kept fixed, $\Omega h^2$ actually
grows. Since what are relevant to experiments are rather the
physical observables (masses, branching ratios, etc.) than the
underlying parameters, we take special care in our analysis to
disentangle effects arising from changes in the couplings from
purely kinematic effects.

After a description of the model in section~2 and its
implementation into \micromegas\ in section~3, we present in
section~4 our results for the typical scenarios of neutralino
annihilation: the mixed bino-Higgsino case with annihilation into
$W$ pairs, rapid annihilation near a Higgs resonance, the
bino-like LSP with light sfermions ($t$-channel and coannihilation
regions), and finally the mixed bino-wino scenario with
neutralino-chargino coannihilation. In section~5, we give a
summary and conclusions. For completeness, the Lagrangian for
sparticle interactions in the CPV-MSSM is given in the Appendix.

\section{The MSSM with CP phases}

We consider the general MSSM with parameters defined at the weak
scale. In this framework, the gaugino and Higgsino mass parameters
and the trilinear couplings can have complex phases,
$M_i=|M_i|e^{i \phi_i}$, $\mu=|\mu|e^{i \phi_\mu}$ and
$A_f=|A_f|e^{i \phi_f}$. Not all of these phases are however
relevant to our analysis. In particular, the phase of $M_2$ can
always be rotated away, while the phase of $M_3$ is mostly
relevant for the coloured sector, so we neglect it in this study.
Among the trilinear couplings, $A_t$ has the largest effect on the
Higgs sector, and can thus potentially play an important role for
the relic density. The phase of the selectron coupling $A_e$,
as the phase of all light sfermions, is irrelevant for the relic
density; it has, however, to be taken into account since it
contributes to electric dipole moments (EDMs).

In the following, we explain our notation and conventions.
We hereby basically follow the notation in {\tt CPsuperH} and
use the SUSY Les Houches Accord (SLHA) \cite{Skands:2003cj}. 
We also discuss some of the most relevant couplings. For the 
complete interaction Lagrangian, see the Appendix and 
Ref.~\cite{Lee:2003nt}.\\

\noi {\bf Neutralinos:}
The neutralino mass matrix in the bino--wino--Higgsino basis \\
$\psi_j^0=(-i\lambda ',\,-i\lambda^3,\,\psi_{H_1}^0,\,\psi_{H_2}^0)$ is
\begin{equation}
  {\cal M}_N =
  \left( \begin{array}{cccc}
  M_1 & 0 & -m_Z s_W c_\beta  & m_Z s_W s_\beta \\
  0 & M_2 &  m_Z c_W c_\beta  & -m_Z c_W s_\beta  \\
  -m_Z s_W c_\beta & m_Z c_W c_\beta   & 0 & -\mu \\
   m_Z s_W s_\beta & - m_Z c_W s_\beta & -\mu & 0
  \end{array}\right)
\label{eq:ntmassmat}
\end{equation}
with $s_W=\sin\theta_W$, $c_W=\cos\theta_W$, $s_\beta=\sin\beta$,
$c_\beta=\cos\beta$ and $\tan\beta = v_2/v_1$
($v_{1,2}$ being the vacuum expectation values of the two Higgs
fields $H_{1,2}$).
This matrix is diagonalized by a unitary mixing matrix $N$,
\begin{equation}
  N^*{{\cal M}_N} N^\dagger =
  {\rm diag}(\mnt{1},\,\mnt{2},\,\mnt{3},\,\mnt{4})\,,
\end{equation}
where $\mnt{i}$, $i=1,...,4$, are the (non-negative) masses
of the physical neutralino states with $\mnt{1}<....<\mnt{4}$.
The lightest neutralino is then decomposed as
\beqn
  \nt_1= N_{11}\tilde{B}+ N_{12} \tilde{W} +N_{13}\tilde{H_1}+
  N_{14}\tilde{H_2}
\eeqn
with the bino ($f_B$), wino ($f_W$) and Higgsino ($f_H$) fractions
defined as
\begin{equation}
   f_B=|N_{11}|^2\,,\qquad
   f_W=|N_{12}|^2\,,\qquad
   f_H=|N_{13}|^2+|N_{14}|^2\,.
\end{equation}
The LSP will hence be mostly bino, wino, or Higgsino according to
the smallest mass parameter in Eq.~(\ref{eq:ntmassmat}),
$M_1$, $M_2$, or $\mu$, respectively. \\

\noi {\bf Charginos:}
The chargino mass matrix in SLHA notation,
\footnote{Here note that in CPsuperH, ${\cal M}_C\rightarrow {\cal M}_C^T$.}
\begin{equation}
  {\cal M}_C =
  \left( \begin{array}{cc}
    M_2 &\sqrt 2\, m_W\sin\beta \\
    \sqrt 2\,m_W\cos\beta & \mu
  \end{array}\right) \,.
\end{equation}
is diagonalized by two unitary matrices $U$ and $V$,
\begin{equation}
  U^*{\cal M}_C V^\dagger = {\rm diag}(\mch{1},\,\mch{2})\,,
\end{equation}
with the eigenvalues again being mass-ordered. The chargino sector
will be relevant  mainly because t-channel chargino exchange
dominates the $\nt_1\nt_1\to W^+W^-$ cross-section, which is often
the main annihilation process when the LSP has a sizeable Higgsino
component. The relevant Lagrangian writes
\begin{equation}
  {\cal L}_{{\tilde \chi^0_j}{\tilde \chi^+_i} W^-} = g W_\mu^-
  \overline{\tilde\chi_j^0} \gamma^\mu
  \left( O^L_{ji} P_L + O^R_{ji} P_R \right) \tilde{\chi}_i^+
  + {\rm h.c.}
\end{equation}
with
\begin{equation}
  O_{ji}^L = N_{j2}V_{i1}^*-\frac{1}{\sqrt{2}} N_{j4} V_{i2}^*\,,\quad
  O_{ji}^R = N_{j2}^*U_{i1}+\frac{1}{\sqrt{2}} N_{j3}^* U_{i2}\,.
\label{eq:ntchw}
\end{equation}
These expressions are the same as for the real MSSM and show that
the $\nt_1\ti\chi^\pm_iW^\mp$ coupling vanishes in case of a pure
bino LSP. Note that this coupling also enters the chargino
coannihilation processes into gauge boson pairs which are
important in some specific scenarios.
Nonzero phases will affect directly the masses of the charginos
and neutralinos as well as the couplings through shifts in the
mixing matrices. \\


\noi {\bf Sfermions:}
Ignoring intergenerational mixing, the
sfermion mass matrices are
\begin{equation}
  {\cal M}_{\ti f}^2 =
  \left( \begin{array}{cc}  \msf{L}^2 &  h_f^* a_f^* \\
                            h_f a_f   & \msf{R}^2
  \end{array}\right)
\label{eq:msfmat}
\end{equation}
with
\begin{eqnarray}
  \msf{L}^2 &=& M^2_{\{\ti Q,\ti L\}}
    + m_Z^2 \cos 2\beta\,(I_{3L}^f - e_f\sin^2\theta_W) + m_f^2, \\[2mm]
  \msf{R}^2 &=& M^2_{\{\ti U,\ti D,\ti R\}}
    + e_f\,m_Z^2 \cos 2\beta\,\sin^2\theta_W + m_f^2,
\end{eqnarray}
and
\begin{eqnarray}
  a_t &=& (A_t\,v_2 - \mu^* v_1)/\sqrt{2} \,,\\
  a_{b,\tau} &=& (A_{b,\tau}\,v_1 - \mu^* v_2)/\sqrt{2} \,.
  \label{eq:aq}
\end{eqnarray}
for $\ti f=\ti t,\,\ti b,\,\ti\tau$. Here
$M_{\{\ti Q,\ti L\}}$, $M_{\{\ti U,\ti D,\ti R\}}$ are the left and
right sfermion mass parameters;
$m_f$, $e_f$ and $I_{3}^f$
are the mass, electric charge and the third component of the weak
isospin of the partner fermion, respectively, and
$h_f$ denote the Yukawa couplings,
$h_t = \frac{\sqrt{2} m_t}{v\cos\beta}$ and
$h_{b,\tau} = \frac{\sqrt{2} m_{b,\tau}}{v\sin\beta}$ at tree level.
Here, $v^2=v_1^2+v_2^2$.
The mass matrix ${\cal M}_{\tilde f}^2$ is diagonalized by a
unitary matrix $R^{\tilde f}$ such that
\begin{equation}
   R^{\tilde f\,\dagger}\,{\cal M}_{\tilde f}^2\, R^{\tilde f} =
   {\rm diag}(m_{\tilde f_1}^2,\,m_{\tilde f_2}^2)
\end{equation}
with $m_{\tilde f_1}\le m_{\tilde f_2}$
and $(\tilde f_L^{},\,\tilde f_R^{})^T =
     R^{\tilde f} (\tilde f_1^{},\,\tilde f_2^{})^T$.
The relevant sfermion interactions are given in the Appendix. \\

\noi {\bf Neutral Higgs bosons:}
In the CP-conserving MSSM, the Higgs sector consists of two CP-even
states $h^0$, $H^0$ and one CP-odd state $A^0$.
Allowing for CP-violating phases induces a mixing between
these three states through loop corrections.
The resulting mass eigenstates $h_1,h_2,h_3$
(with $m_{h_1}<m_{h_2}<m_{h_3}$ by convention) are no
longer eigenstates of CP. The Higgs mixing matrix is defined by
\footnote{In CPsuperH, the Higgs mixing matrix is denoted $O_{ai}$.}
\begin{equation}
   (\phi_1,\phi_2,a)^T_a =H_{ai} (h_1,h_2,h_3)_i^T.
\end{equation}
Because of the mixing between the neutral scalar and pseudoscalar states,
it is preferable to use the charged Higgs mass, $m_{H^+}$, as an
independent parameter.  In what follows we will mainly be
concerned with the couplings of the lightest neutralino to Higgs
bosons that govern the neutralino annihilation cross section via
Higgs exchange. The Lagrangian for such interactions writes
\begin{equation}
  {\cal L}_{{\tilde \chi^0_1}{\tilde \chi^0_1} h_i}=-
  \frac{g}{2}{\sum_{i=1}^3}\,\overline{\tilde\chi_1^0} (
  g^{S}_{h_i{\tilde \chi^0_1}{\tilde \chi^0_1}}+ i \gamma_5
  g^{P}_{h_i{\tilde \chi^0_1}{\tilde \chi^0_1}} )\tilde{\chi}_1^0 h_i
\end{equation}
with  the scalar and pseudoscalar couplings corresponding to
the real and imaginary part of the same expression:
\begin{eqnarray}
\label{eq:gs}
  g^{S}_{h_i{\tilde \chi^0_1}{\tilde \chi^0_1}}
    &=& Re\left[(N_{12}^*-t_W N_{11}^*)
        \left(H_{1i}N_{13}^*-H_{2i}N_{14}^*-i H_{3i}
        (s_\beta N_{13}^* - c_\beta N_{14}^*)\right) \right],\\
  g^{P}_{h_i{\tilde \chi^0_1}{\tilde \chi^0_1}}
    &=& -Im\left[(N_{12}^*-t_W N_{11}^*)
        \left(H_{1i}N_{13}^*-H_{2i}N_{14}^*-i H_{3i}
        (s_\beta N_{13}^* - c_\beta N_{14}^*)\right) \right].
\end{eqnarray}
For real parameters, one recovers the $(h^0\!,H^0\!,A^0)$ couplings
by taking for the pseudoscalar, $A^0$, $H_{33}=1$ and all other
$H_{ij}=0$, while for the scalar Higgs couplings,
the only nonzero elements are $H_{11}=-H_{22}=-\sin\alpha$
and $H_{21}=H_{12}=\cos\alpha$.
The LSP couplings to neutral Higgs bosons will clearly be affected
both by phases in the neutralino sector 
which modify the neutralino mixing, as well as by phases which
induce scalar-pseudoscalar mixing in the Higgs sector.
Indeed in the MSSM the Higgs CP mixing is induced by loops involving
top squarks and is proportional to
$Im(A_t\mu)/(m^2_{{\tilde{t}}_2}-m^2_{{\tilde{t}}_1})$
\cite{Choi:2002zp}.
Thus a large mixing is expected when $Im(A_t\mu)$ is comparable
to the stop masses-squared. Note that
the masses of the physical Higgs bosons also depend on the phases
of $A_t$ and $\mu$. In particular, a large mass splitting between the
heavy states is found for large values of $\mu A_t$.
For a more detailed discussion of Higgs masses and couplings
in the CPV-MSSM, we refer to \cite{Lee:2003nt}.

\section{Relic density computation and
     implementation of the CPV-MSSM into \micro}

The computation of the relic density of dark matter is by now
standard~\cite{Gondolo:1990dk,Gondolo:1998ah}. It relies on
solving the evolution equation for the abundance, defined as the
number density divided by the entropy density,
\begin{equation}
  \frac{dY}{dT} = \sqrt{\frac{\pi  g_*(T) }{45}} M_p \,
                  \langle\sigma v\rangle
                  \big(Y(T)^2-Y_{eq}(T)^2\big)
                = A(T)\,\big(Y(T)^2-Y_{eq}(T)^2\big)\,,
  \label{dydt}
\end{equation}
where $g_{*}$ is an effective number of degrees of freedom
\cite{Gondolo:1990dk}, $M_p$ is the Planck mass and $Y_{eq}(T)$
the thermal equilibrium abundance, and $\sigv$ the
relativistic thermally averaged annihilation cross-section of
superparticles summed over all channels. It is given by
\begin{equation}
       \sigv =  \frac{ \sum\limits_{i,j}g_i g_j
\int\limits_{(m_i+m_j)^2} ds\sqrt{s}
K_1(\sqrt{s}/T)\, p_{ij}^2\sum\limits_{k,l}\sigma_{ij;kl}(s)}
                         {2T\big(\sum\limits_i g_i m_i^2 K_2(m_i/T)\big)^2
}\,,
\label{sigmav}
\end{equation}
with $g_i$  the number of degrees of freedom,  $\sigma_{ij;kl}$
the total cross section for annihilation of a pair of
supersymmetric particles with masses $m_i$, $m_j$ into some
Standard Model particles ($k,l$), and $p_{ij}\;(\sqrt{s})$  the
momentum (total energy) of the incoming particles in their
center-of-mass frame.
Integrating Eq.~(\ref{dydt}) from $T=\infty$ to $T=T_0$ gives
the present day abundance $Y(T_0)$
needed in the evaluation of the relic density,
\beqn
  \label{omegah} \Omega_{\rm LSP} h^2=
  \frac{s(T_0)h^2}{\rho_{cr}} M_{\rm LSP}Y(T_0)=
  2.742 \times 10^8 \frac{M_{\rm LSP}}{\rm GeV} Y(T_0)
\eeqn
where $s(T_0)$ is the entropy density at present time and
$h$ the normalized Hubble constant. The present-day energy density
is then simply expressed as $\rho_{\rm LSP}=10.54 \Omega h^2
(\rm GeV/m^3)$.

To compute the relic density, \micro\ solves the equation for the
abundance, Eq.~(\ref{dydt}), numerically without any approximation.
In addition, \micro\ also estimates the relative contribution of
each individual annihilation or coannihilation channel to the
relic density. For this specific purpose, the freeze-out
approximation is used: 
below the freeze-out temperature $T_f$,
$Y_{eq}\ll Y$,  and Eq.~(\ref{dydt}) can be integrated
\begin{equation}
     \frac{1}{Y(T_0)}= \frac{1}{Y(T_f)} + \int\limits_{T_0}^{T_f}A(T)dT\,.
\label{fout1}
\end{equation}
It turns out that $M_{\rm LSP}/T_f\approx 25$ 
(for more details see~\cite{Gondolo:1990dk,Belanger:2004yn}).
In this case, the first term in Eq.~(\ref{fout1}) is
suppressed, and one can approximate $1/\Omega$
by a sum over the different annihilation channels:

\begin{equation}
\frac{1}{\Omega h^2} \approx \sum \limits_{i,j,k,l}
\frac{1}{\omega_{ij;kl}h^2}
\end{equation}
where
\begin{equation}
   \omega_{ij;kl} = \frac{s(T_0)M_{\rm LSP}}{\rho_{cr}}
   \int\limits_{T_0}^{T_f}\sqrt{\frac{\pi g_*(T)}{45}} M_p
   \frac{ g_i g_j  \int ds\sqrt{s}
          K_1(\sqrt{s}/T) p_{ij}^2 \sigma_{ij;kl}(s)}
        {2T\big(\sum\limits_{i'} g_{i'} m_{i'}^2 K_2(m_{i'}/T)\big)^2 } dT \,.
\label{yij}
\end{equation}

Note that this method gives to a good approximation the
contribution of individual channels, although the value of $\Omega
h^2$ is slightly overestimated as compared to the exact value
obtained by solving directly the evolution equation for the
abundance as described above.

To perform this computation for the CPV-MSSM,
we are using \micromegas~\cite{mo2:Allanach:2006fy}, an extension
of {\tt micrOMEGAs1.3}~\cite{Belanger:2004yn,Belanger:2001fz} that
allows the computation of the relic density of dark matter within
generic particle physics models that feature a stable weakly
interacting particle. Within this framework we have implemented 
the CPV-MSSM as follows. Using {\tt LanHEP}~\cite{Semenov:2002jw},
a new CPV-MSSM model file with complex parameters was rebuilt in
the {\tt CalcHEP}~\cite{Pukhov:2004ca} notation, thus specifying
all relevant Feynman rules. For the Higgs sector, an effective
potential was written in order to include in a consistent way
higher-order effects \cite{Pilaftsis:1999qt}. Masses, mixing
matrices and parameters of the effective potential are read
directly from {\tt CPsuperH}~\cite{Lee:2003nt}, together with
masses and mixing matrices of neutralinos, charginos and third
generation sfermions. The masses of the first two generations of
sfermions are computed at tree-level from the MSSM input
parameters within \micro. The cross sections for all annihilation
and coannihilation processes are computed automatically with {\tt
CalcHEP}. The standard \micro\ routines are used to calculate the
effective annihilation cross section and the relic density of the
LSP. 
This CPV-MSSM version of {\tt micrOMEGAs} has first been presented 
in~\cite{omcp:Allanach:2006fy}.

\subsubsection*{EDM constraint}
We have also implemented the constraints originating from the
electric dipole moment  of the electron (eEDM), $d_e<2.2\times
10^{-27} {\rm e\,cm}$~\cite{Hagiwara:2002fs}. In the MSSM, for
small to intermediate values of $\tan\beta$, the dominant
contribution to $d_e$ originates from one-loop chargino/sneutrino
and neutralino/selectron exchange
diagrams~\cite{Ibrahim:1998je,Choi:2004rf}. At one loop, the
expression for $d_e$ depends on the complex parameters in the
neutralino/chargino sector ($M_1,\mu$) as well as  on the
trilinear coupling of the electron, $A_e$.
Note that $A_e$ is suppressed by a factor proportional to
$m_e$ and is completely negligible for cross sections
calculations. It is, however, relevant for the eEDM since $d_e$ is
itself proportional to the electron mass.  The eEDM features a
strong dependence on $\mu$ because the necessary helicity flip
originates from  the coupling of the electron to the Higgsino,
which is proportional to $\mu$. The one-loop contributions to the
eEDM basically restrict $\phi_\mu$ to $|\sin\phi_\mu| \sim {\cal
O}(10^{-2})$ with sfermions at the TeV scale. The restrictions on
the phases $\phi_1$ and $\phi_e$ are more modest.
At one loop, the eEDM constraint can be most easily evaded by
raising the masses of the sfermions of the first generation, say
to 10~TeV. However, for such high sfermion masses, two-loop
contributions~\cite{Chang:1998uc,Pilaftsis:1999td} can also become
important, especially if $A_{t,b}$ and $\mu$ are also in the
TeV range.  Two-loop contributions are further enhanced for small
values of the charged Higgs mass and for large
$\tan\beta$~\cite{Demir:2003js}. In the scenarios which we
consider here, we will sometimes have to rely on a cancellation
between one- and two-loop contributions to have an acceptable
value for the eEDM.

\section{Results}

We now turn to the numerical analysis and present results for
various scenarios for which the relic density is in agreeement
with WMAP. We impose GUT relations for the gaugino masses,
$M_1:M_2:M_3\approx 1:2:6$, unless mentioned otherwise. In order
to satisfy the eEDM constraint, we assume in most cases that the
sfermions of the first and second generation are heavy, $m_{\tilde
f_{L,R}}=10$~TeV, allowing only the third generation to be at the
TeV scale. Unless specific values are specified, we take
$M_S\equiv M_{\ti Q_3}=M_{\ti U_3}=M_{\ti D_3}=M_{\ti L_3}=M_{\ti
R_3}$. For the trilinear couplings, we keep $A_t$ as a free
parameter, assuming $|A_f|=1$~TeV for all others. This is
justified as the $A_f$ with $f\not=t$ have only a very mild effect
on the neutralino cross-sections. However, one has to keep track
of $A_e$ because the electron EDM depends on the phase in the
slepton sector. For simplicity we consider a common phase $\phi_l$
for all trilinear slepton couplings. In general, the parameters that 
will be allowed to vary hence are
\begin{equation}
  |M_1|,\; |\mu|,\; \tan\beta,\; m_{H^+},\; |A_t|,\; M_{S},\;
  \phi_1,\; \phi_\mu,\; \phi_t,\; \phi_l \,.
\end{equation}
As mentioned above, the eEDM will constrain $\phi_\mu$ to $\sim 0$
or $180^\circ$.
From now on we drop the $||$ for simplicity of notation,
i.e.\ $|M_1|e^{i\phi_1}\to M_1 e^{i\phi_1}$,  etc.
So when specifying the value of a complex parameter,
we implicetly mean its absolute value.

The cross sections for the annihilation and coannihilation
processes will depend on phases, and so will the
thermally-averaged cross section $\sigv$, Eq.~(\ref{sigmav}).
Part of this is due to changes in the physical masses,
leading to huge variations in the relic density
especially when coannihilation processes are
important or when annihilation occurs near a resonance.
We will therefore take special care to disentangle the effects
from kinematics and couplings.
Indeed, as we will see, in many cases a large part of the phase
dependence can be explained by changes in the masses of the
involved particles.
However, in some cases disentangling the kinematic effects will
also lead to an enhancement of the phase dependence.

The scenarios which we consider are the typical scenarios for
neutralino annihilation:
the mixed bino-Higgsino LSP that annihilates into gauge bosons,
the rapid annihilation through a Higgs resonance,
coannihilation with third generation sfermions, and
finally a scenario with a mixed bino-wino LSP.
The case of $t$-channel exchange of light sfermions is discussed
together with the sfermion coannihilation.

\subsection{The mixed bino-Higgsino LSP}
\label{sec:Higgsino}

We start with the case that all scalars except the light Higgs are
heavy, $M_S=m_{H^+}=1$~TeV. In this scenario we do not expect a
dependence of the relic density on the phase of the slepton
sector; we therefore set $\phi_l=\phi_t$. In the real MSSM,
a bino-like LSP with a mass of the order of 100~GeV needs a
Higgsino admixture of roughly 25\%--30\% for its relic density to
be within the WMAP range
\cite{Belanger:2004hk,Arkani-Hamed:2006mb, Masiero:2004ft}. In
terms of fundamental MSSM parameters this means $M_1\approx \mu$.
The main annihilation mechanisms then are $\nt_1\nt_1\to WW$ and
$ZZ$ through $t$-channel chargino and neutralino exchange, as well
as $\nt_1\nt_1\to t\bar t$ when kinematically allowed. The latter
proceeds through $s$-channel $Z$ or $h_1$ exchange. The LSP
Higgsino fraction determines the size of the annihilation
cross-section because it directly enters the $\nt_1\chpm_iW^\mp$
and $\nt_1\nt_j Z$ vertices.

\begin{figure}[t]
  \centerline{\epsfig{file=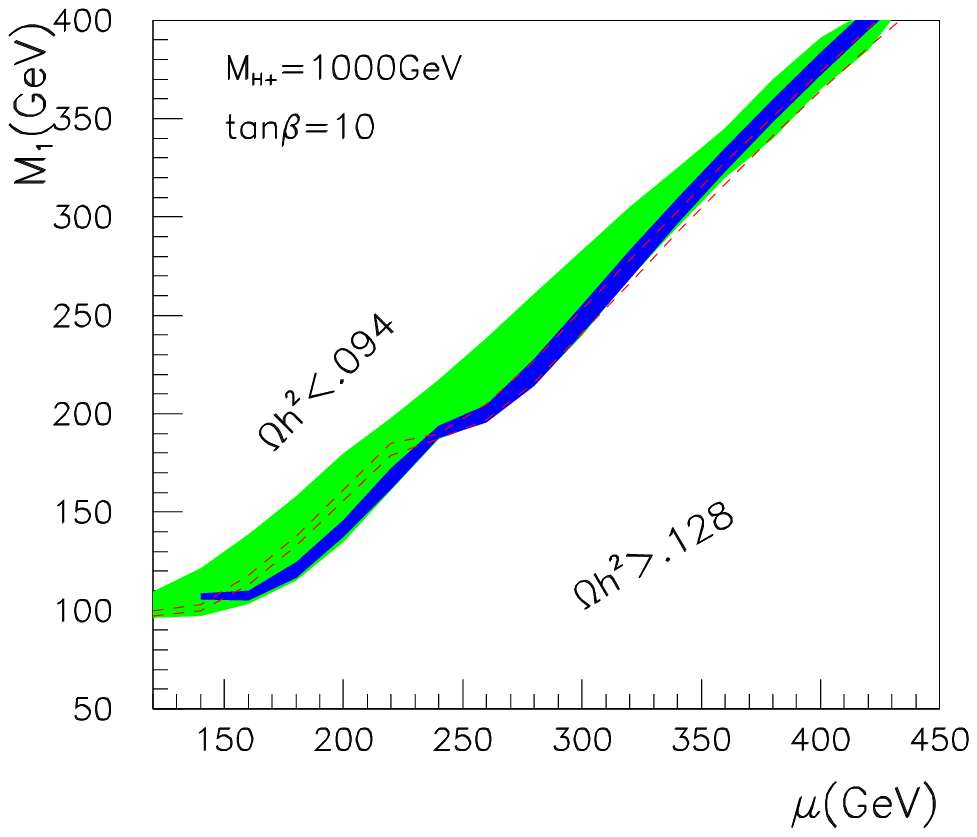, width=7.2cm}\qquad
              \epsfig{file=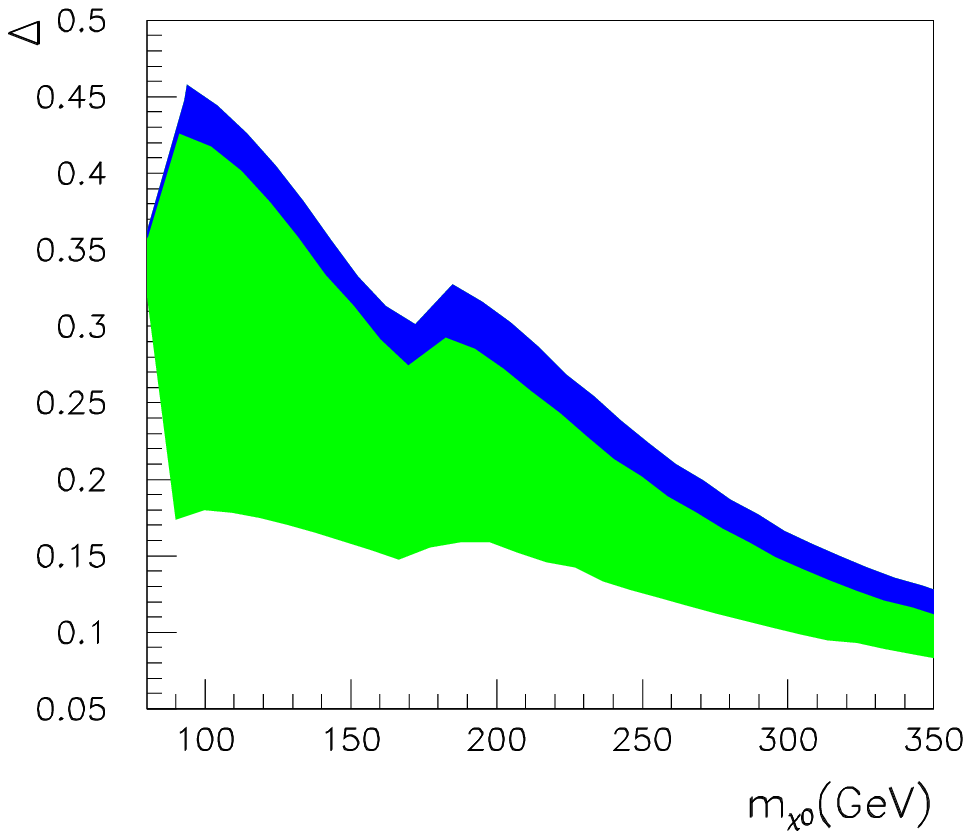, width=7.2cm}}
  \ablabels{-2}{78}{62}
  \caption{(a) The $2\sigma$ WMAP bands in the $M_1$--$\mu$
           plane for $\tan\beta=10$, $m_{H^+}=M_S=A_t=1$~TeV,
           for all phases zero (blue/dark grey band),
           for $\phi_\mu=180^\circ$ (or $\mu<0$) and all other phases zero
           (dashed red lines)
           and for arbitrary phases (green/light grey band).
           (b) The corresponding relative mass difference
               $\Delta=\Delta m_{\nt_1\ch_1}/\mnt{1}$
               as function of $\mnt{1}$ in the $2\sigma$ WMAP band
               for all phases zero (blue/dark grey band)
               and for arbitrary phases (green/light grey band).}
\label{fig:m1mumh1000}
\end{figure}

We perform a scan in the $M_1$--$\mu$ plane and display in
Fig.~\ref{fig:m1mumh1000}a the region where the relic density
is in agreement with the $2\sigma$ WMAP bound, Eq.~(\ref{eq:wmap}).
In the real MSSM, when all phases are zero, only the narrow
blue (dark grey) band is allowed. This band shifts slightly for
negative values of $\mu$ ($\phi_\mu=180^\circ$).
The onset of the $t\bar{t}$ annihilation channel appears as a kink.
When allowing all phases to vary arbitrarily, keeping only those
points for which all constraints are satisfied for at least one
combination of phases, the allowed band becomes much wider, see
the green (light grey) band in Fig.~\ref{fig:m1mumh1000}a.
For a given $M_1$, the allowed
range for $\mu$ increases roughly from $\delta\mu\sim 10$~GeV
to $\delta\mu\sim 50$~GeV. Since the eEDM constraint results in
$\phi_\mu$ close to zero or $180^\circ$, 
this is actually mostly due to $\phi_1$. In fact the left boundary
of the green band roughly corresponds to the contour of
$\Omega h^2=0.0945$ for $\phi_1=\phi_\mu=180^\circ$.

\begin{figure}[t]
  \centerline{\epsfig{file=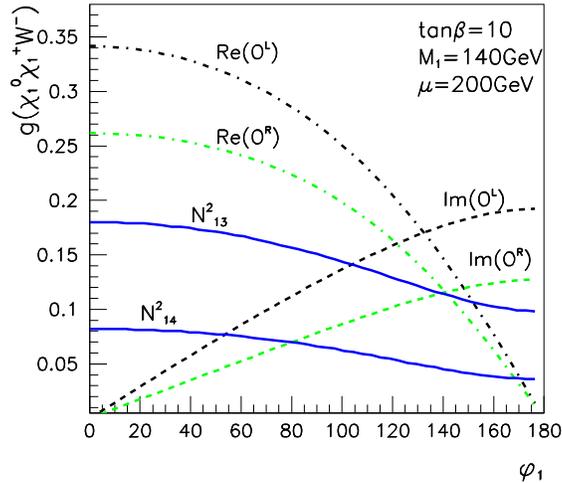, width=8cm}}
  \caption{The $\nt_1\ch_1W^-$ couplings, $O^{L,R}_{11}$,
  Eq.~(\ref{eq:ntchw}), as function of $\phi_1$, for $M_1=140$~GeV,
  $\mu=200$~GeV, $\phi_\mu=\phi_t=0$, and the other parameters as
  in Fig.~\ref{fig:m1mumh1000}. The Higgsino fractions of the LSP,
  $|N_{13}|^2$ and $|N_{14}|^2$, are also displayed.}
\label{fig:gntchw}
\end{figure}

In Fig.~\ref{fig:m1mumh1000}b, we show the relative mass difference
between the lightest chargino and the LSP,
$\Delta m_{\nt_1\ch_1}/\mnt{1}=(\mch{1}-\mnt{1})/\mnt{1}$,
in the WMAP-allowed bands of Fig.~\ref{fig:m1mumh1000}a.
As a general feature, $\Delta m_{\nt_1\ch_1}/\mnt{1}$ decreases with
increasing $\mnt{1}$ because the cross-sections for the pair-annihilation
channels decrease, and coannihilation channels are needed in addition 
to maintain $\Omega h^2\sim 0.1$.
In the CPV-MSSM, however, for a given $\mnt{1}$ much smaller
mass differences can be in agreement with the WMAP bound
than in the CP-conserving case.
This is because, as we will discuss in more details below,
the $\nt_1\chpm_1W^\mp$ couplings decrease with increasing $\phi_1$,
so that additional contributions of coannihilation
channels are required to maintain compatibility with WMAP.
The phase dependence of the $\nt_1\chpm_1W^\mp$ couplings
is shown in Fig.~\ref{fig:gntchw} for $M_1=140$~GeV and $\mu=200$~GeV.

\subsubsection{Below the \boldmath $t\bar t$ threshold}

\begin{figure}[p]
  \centerline{\epsfig{file=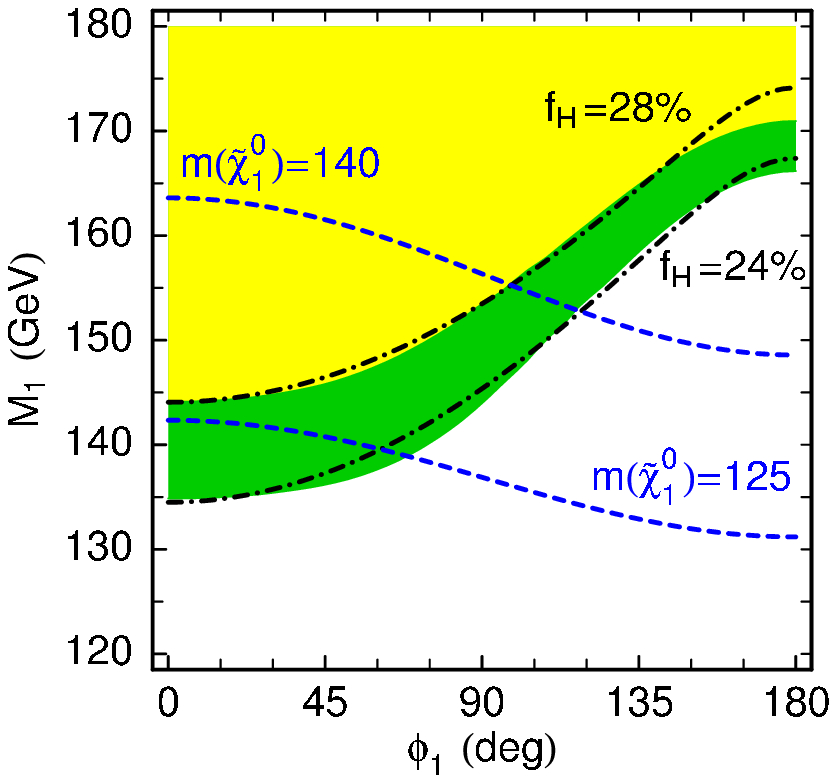, width=7cm}\qquad
              \epsfig{file=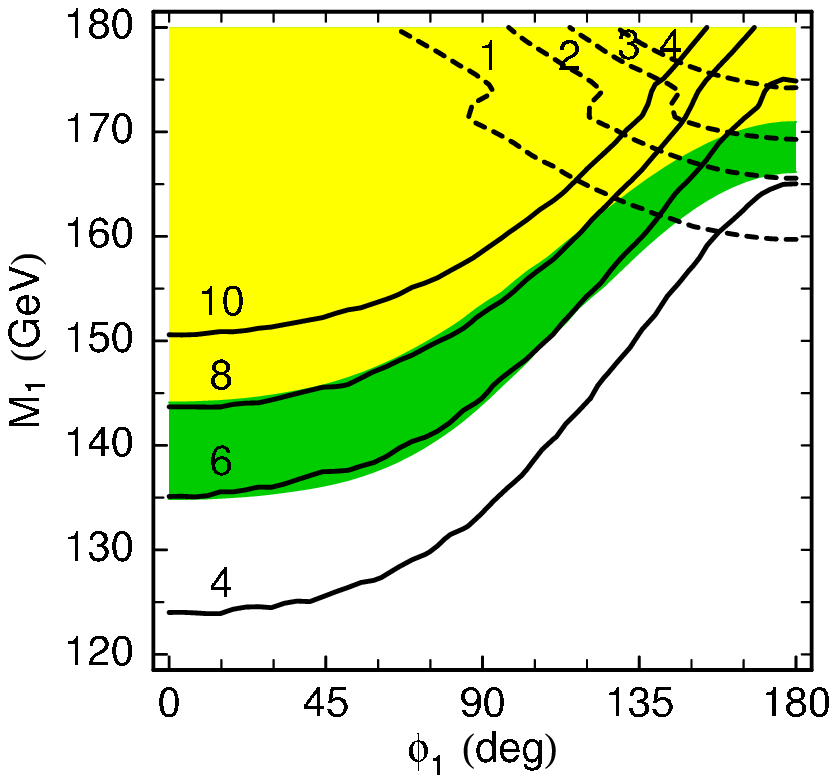, width=7cm}}
  \ablabels{2}{81}{70}
  \caption{The $2\sigma$ WMAP band (green/dark grey) in the
  $M_1$--$\phi_1$ plane for $\mu=200$~GeV, $\tan\beta=10$,
  $m_{H^+}=1$~TeV, $A_t=1.2$~TeV, $\phi_t=\phi_\mu=0$.
  In the yellow (light grey) region $\Omega h^2$ is below the
  WMAP bound, and in the white region it is too large.
  Superimposed are in (a) contours of constant LSP mass,
  $\mnt{1}=125$ and 140~GeV (dashed lines) and contours of constant
  LSP Higgsino fraction, $f_H=24\%$ and $28\%$ (dash-dotted lines).
  In (b) contours of constant $(\omega_{ij;kl} h^2)^{-1}$, Eq.~(\ref{yij}),
  are shown;
  the full lines are for $\nt_1 \nt_1\rightarrow W^+W^-,ZZ$ and the
  dashed lines for $\nt_1\chpm_1$ coannihilation channels.}
\label{fig:m1phi1}
\end{figure}

Figure~\ref{fig:m1phi1} shows the WMAP band in the
$M_1$--$\phi_1$ plane ($\phi_\mu=\phi_t=\phi_l=0$) for
$\mu=200$~GeV and the other parameters as above.
Also shown are contours of constant LSP mass,
contours of constant LSP Higgsino fraction $f_H$,
as well as contours of constant cross-sections for the
main annihilation channels. We can make several observations.
First, the mass of the LSP increases with $\phi_1$.
On the one hand this induces a decrease in the LSP pair-annihilation
cross-sections. On the other hand, since the chargino mass is
independent of $\phi_1$, the NLSP--LSP mass splitting is reduced,
making coannihilation processes with $\ch_1$ (and also $\nt_2$)
more important.
Second, the Higgsino fraction decreases with increasing $\phi_1$.
The left- and right-handed $\nt_1\chpm_1W^\mp$ couplings feature the
same phase dependence, see Fig.~\ref{fig:gntchw}.
The modification of the LSP couplings to gauge bosons leads to a decrease
in the dominant $\nt_1\nt_1\to WW/ZZ$ cross-sections
and thus a higher value for the relic density.
Here note that the phase dependence of $f_H$ (and hence of the couplings)
is much more pronounced than that of the LSP mass.
As a result, in Fig.~\ref{fig:m1phi1}a there is an almost perfect match
between the $2\sigma$ WMAP band and the band of $24\%\le f_H \le 28\%$.
The small deviation at $\phi_1\sim 40^\circ$--$90^\circ$ comes
from the subdominant annihilation into $Zh_1$ and $h_1h_1$.
The larger deviation at $\phi_1\sim 150^\circ$--$180^\circ$
comes from coannihilations, c.f.\ Fig.~\ref{fig:m1phi1}b.

\begin{figure}[p]
  \centerline{\epsfig{file=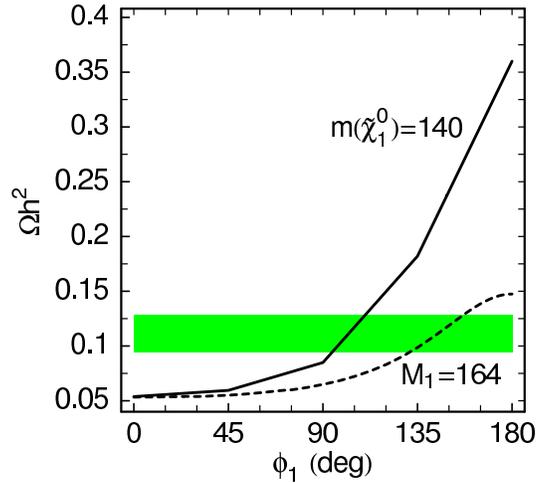, width=7cm}}
  \caption{$\Omega h^2$ as a function of $\phi_1$ for the parameters
  of Fig.~\ref{fig:m1phi1}; the dashed line is for fixed $M_1=164$~GeV,
  while for the full line $M_1$ is adjusted such that $\mnt{1}=140$~GeV.
  The green (grey) band shows the $2\sigma$ WMAP-allowed range.}
\label{fig:mlsp140}
\end{figure}

These features also explain why in Fig.~\ref{fig:m1mumh1000}a
nonzero phases only extend the WMAP-allowed range into the region
where $\Omega h^2<0.094$ in the real case, and not into the one
where $\Omega h^2$ is too large. Moreover, note that the phase
dependence is large when annihilation into gauge bosons dominates,
but weakened by contributions from coannihilation processes.

To isolate the effect that comes solely from modifications in
couplings, we display in Fig.~\ref{fig:mlsp140} the variation of
$\Omega h^2$ as function of $\phi_1$ for constant
$\mnt{1}=140$~GeV. For comparison, the variation of $\Omega h^2$
for $M_1=164$~GeV (leading to $\mnt{1}\simeq 140$~GeV at
$\phi_1=0$) is also shown. In this example, $\Omega h^2$ varies by
a factor of 3 for constant $M_1$, and by a factor of 7 for
constant LSP mass. Overall, we find a phase dependence in $\Omega
h^2$ of almost an order of magnitude for constant LSP mass. We
emphasize that in this case of a mixed bino-Higgsino LSP, the
dependence of masses and couplings on $\phi_1$ work against each
other, so that taking out the kinematic effects actually enhances
the variation of $\Omega h^2$.

\begin{figure}[p]
  \centerline{\epsfig{file=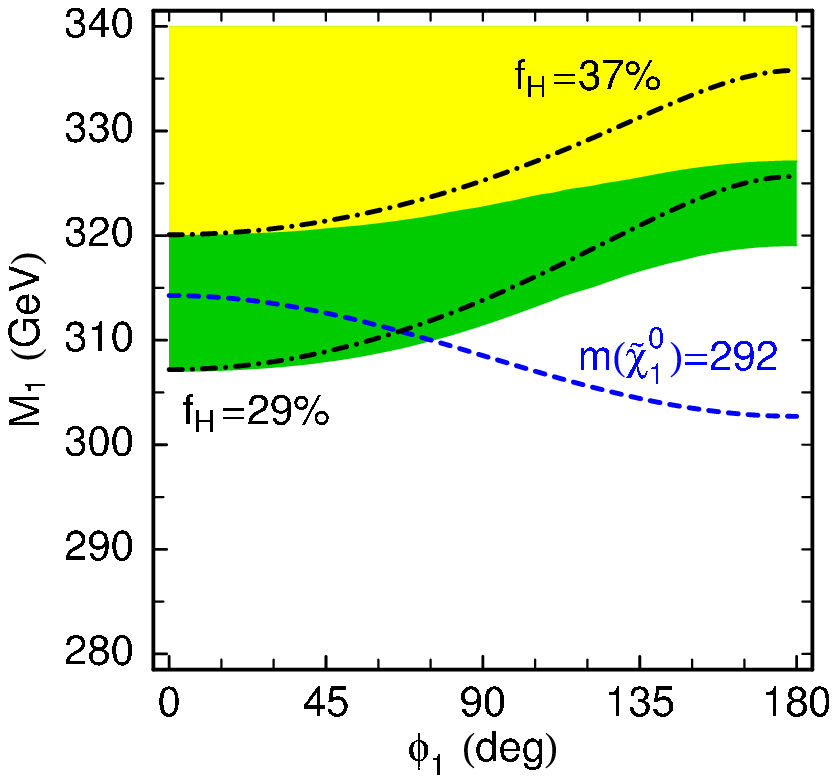, width=7cm}\qquad
              \epsfig{file=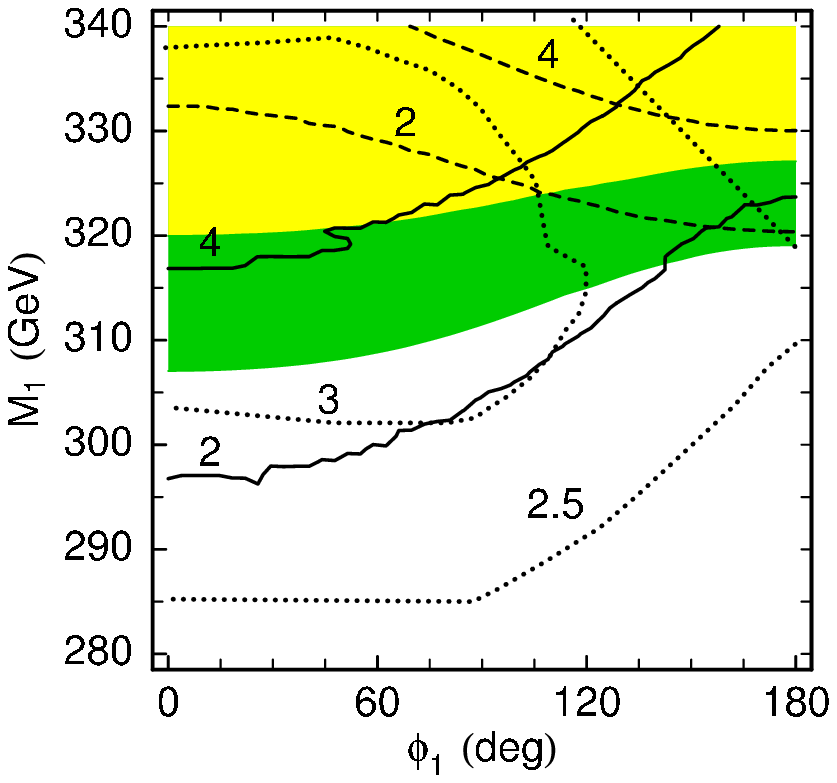, width=7cm}}
  \ablabels{2}{81}{70}
  \caption{The $2\sigma$ WMAP band (green/dark grey) in the
  $M_1$--$\phi_1$ plane for $\mu=350$~GeV, $\tan\beta=10$,
  $m_{H^+}=1$~TeV, $A_t=1.2$~TeV, $\phi_t=\phi_\mu=0$.
  In the yellow (light grey) region, $\Omega h^2<0.0945$.
  Superimposed are in (a) a contour of constant LSP mass
  $\mnt{1}=292$~GeV (dashed line) and contours of constant
  LSP Higgsino fraction $f_H=29\%$ and $37\%$ (dash-dotted lines),
  and in (b) contours of constant $(\omega_{ij;kl} h^2)^{-1}$, 
  full lines: $\nt_1\nt_1\ra WW/ZZ$, dashed lines:
  coannihilation with $\ch_1$ and $\nt_2$, dotted lines:
  $\nt_1\nt_1\ra t\bar t$.}
\label{fig:m1phi1mu350}
\end{figure}
\begin{figure}[p]
  \centerline{\epsfig{file=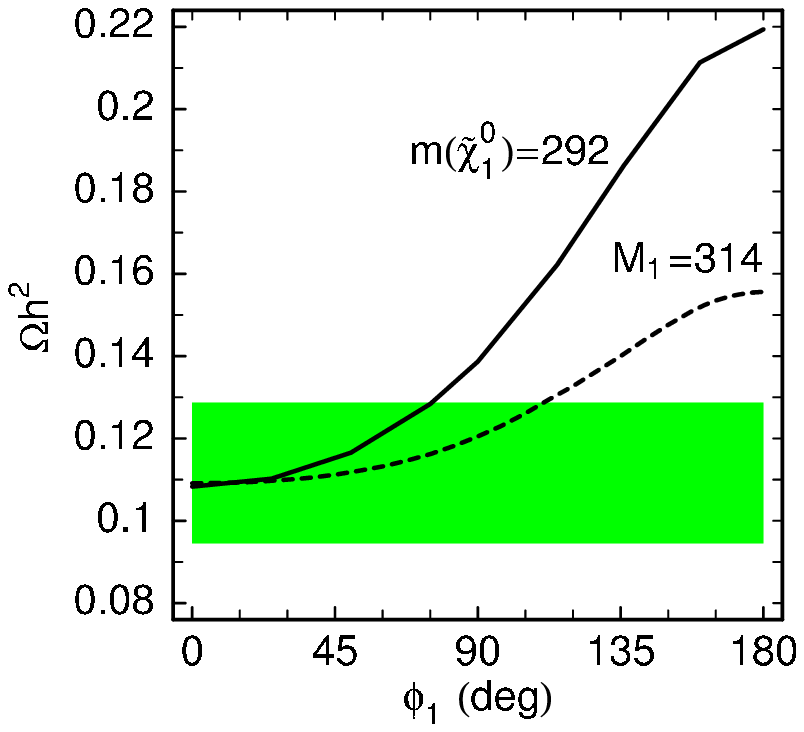, width=7cm}}
  \caption{$\Omega h^2$ as a function of $\phi_1$ for the parameters
  of Fig.~\ref{fig:m1phi1mu350}; the dashed line is for fixed $M_1=314$~GeV,
  while for the full line $M_1$ is adjusted such that $\mnt{1}=292$~GeV.
  The green (grey) band shows the $2\sigma$ WMAP-allowed range.}
\label{fig:mlsp292}
\end{figure}

\subsubsection{Above the \boldmath $t\bar t$ threshold}

In the parameter range of Figs.~\ref{fig:m1phi1} and
\ref{fig:mlsp140} one is always below the $t\bar t$ threshold. We
therefore consider in the next example a higher value of $\mu$,
such that $\nt_1\nt_1\to t\bar t$ contributes in the WMAP-allowed
region. Analogous to Fig.~\ref{fig:m1phi1},
Fig.~\ref{fig:m1phi1mu350} shows the WMAP band in the
$M_1$--$\phi_1$ plane for $\mu=350$~GeV together with contours of
constant LSP mass, Higgsino fraction, and main (co)annihilation
cross-sections. As can be seen in Fig.~\ref{fig:m1phi1mu350}b,
$\nt_1\nt_1\to WW/ZZ$ and $\nt_1\nt_1\to t\bar t$ are of comparable 
importance, each contributing about 40\%--50\% to the total
annihilation cross-section. The top-pair channel, proceeding
through $s$-channel $Z$ or $h_1$ exchange, shows a milder phase
dependence ($s$-channel $h_{2,3}$ and $t$-channel stop exchange
are negligible for this choice of parameters). Were it not for
coannihilations, the lines of constant $\Omega h^2$ would again
follow the lines of constant $f_H$ in Fig.~\ref{fig:m1phi1mu350}a.
However, since we are now dealing with a much heavier LSP, we also
need a larger Higgsino fraction to obtain the right relic density.
At $\phi_1=0$, this means $f_H\simeq 29\%$--37\%. This means in
turn a smaller difference between $M_1$ and $\mu$ and hence a
smaller $\nt_1$--$\ch_1$ mass difference in the WMAP-allowed band
as compared to the previous case. Therefore $\nt_1\ch_1$ and
$\nt_1\nt_2$ coannihilations are relatively more important. Since
their cross-sections show the opposite phase dependence as those
of the pair-annihilations, and since they are mainly determined by
the mass difference, the overall phase dependence of the
WMAP-allowed band is much weakened.

In Fig.~\ref{fig:mlsp292} we show the variation of $\Omega h^2$ as
function of $\phi_1$ for constant $M_1=314$~GeV and for
constant $\mnt{1}=292$~GeV. We see that, taking out kinematic effects,
for a relatively heavy bino-Higgsino LSP the phase dependence of
$\Omega h^2$ is about a factor of 2.

\subsubsection{Lowering \boldmath $m_{H^+}$ to 500~GeV}

Still keeping the sfermions heavy, we next lower the charged Higgs
mass to $m_{H^+}=500$~GeV. This leads to rapid annihilation
via $s$-channel Higgs exchange when
$\mnt{1}\approx m_{h_i}/2\approx 250$~GeV.
In this case only a very small coupling  of the LSP to one of the
Higgses ($h_2,\,h_3$) is necessary, so $\mu$ can be large and the LSP
still annihilates efficiently even if it is dominantly bino.

\begin{figure}
  \centerline{\epsfig{file=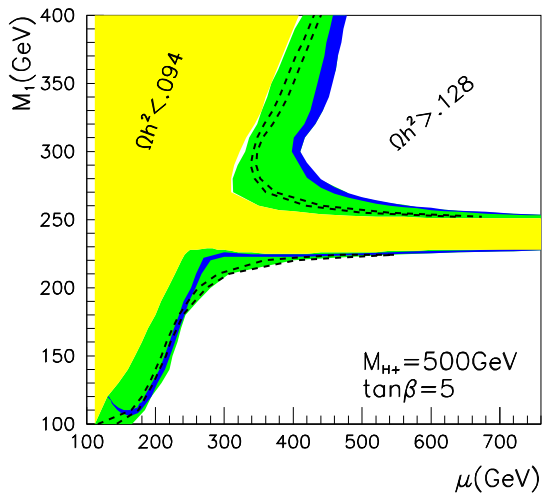, width=8cm}}
  \caption{The $2\sigma$ WMAP bands in the $M_1$--$\mu$ plane
  for $m_{H^+}=500$~GeV, $A_t=1.2$~TeV and $\tan\beta=5$ for all phases zero
  (blue/black bands) and for arbitrary phases (green/light grey bands).
  The other parameters are as in Fig.~\ref{fig:m1mumh1000}. In the yellow region, $\Omega h^2<0.094$.
  The dashed lines show contours of $\Omega h^2=0.094,0.128$ when $\phi_\mu=180^\circ, \phi_1=0$.}
\label{fig:mh500}
\end{figure}

In Fig.~\ref{fig:mh500} we display the WMAP-allowed regions in the
$M_1$--$\mu$ plane for both the real MSSM and the CPV-MSSM.
As before, in the CPV-MSSM, the allowed region corresponds to the
points in the $M_1$--$\mu$ plane for which there exists at least one
choice of $\phi_1$, $\phi_\mu$, $\phi_t$, $\phi_l$ for which all
constraints are satisfied.
One clearly sees the two very narrow bands of the so-called Higgs
funnel at $\mnt{1}\approx m_{H^+}/2$ and large $\mu$.  The impact
of the phases on the relic density in the funnel region will be
discussed in the next section.
Outside this region, i.e.\ for small $\mu$, we observe as before
a significant widening of the band consistent with WMAP for
nonzero phases.
Furthermore, for $\phi_\mu\not=0$, lower values of $\mu$ give
$\mch{1}>103.5$~GeV, consistent with the LEP constraint~\cite{LEPSUSY}
on charginos.
This widening of the WMAP band is again due to shifts in couplings
and masses as we have discussed above. 
Again, the widening is into the region where $\Omega h^2<0.0945$ 
in the CP-conserving case. 
Here note that for $M_1=200$~GeV the extra allowed region to the right of the
blue band corresponds to $\phi_\mu=180^\circ$.

The shifts in masses are specially relevant
in the region around the Higgs funnel. For example at
$M_1=270$~GeV, the band of allowed values for $\mu$ increases from
15 to 160~GeV when allowing for arbitrary phases. Here one is
still close enough to the resonances to have dominant annihilation
through Higgs exchange, and small changes in the $h_{2,3}$ masses
have a large effect on $\Omega h^2$. Furthermore, because the
couplings of the LSP to the heavy Higgses can be suppressed, in
the CPV-MSSM the LSP can have a much larger Higgsino component as
compared to the real case and still be in
agreement with WMAP.

\begin{figure}
  \centerline{\epsfig{file=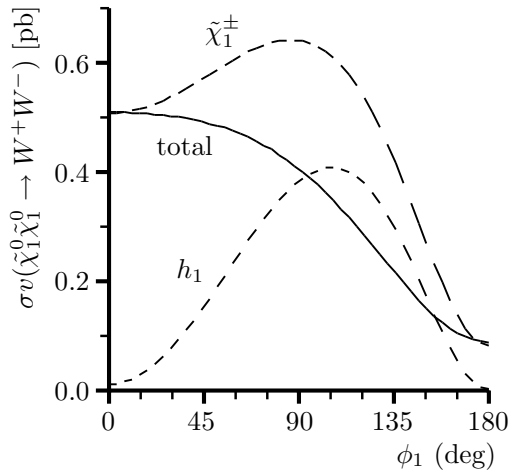, width=7cm}}
  \caption{ Cross section times relative velocity,
  $\sigma v(\nt_1\nt_1\to W^+W^-)$, as function of $\phi_1$ for
  $p=10$~GeV, $M_1=220$~GeV, $\mu=260$~GeV,
  $A_t=1$~TeV and the other parameters as in Fig.~\ref{fig:mh500}.
  The long-dashed line is the $\chpm_1$ exchange, the short-dashed line
  the $h_1$ exchange, and the full line the total contribution.
  Here $p$ is the center-of-mass momentum and $v=2p/\mnt{1}$.
  }
\label{fig:sigvWW}
\end{figure}

A priori one could think that in this region, where several
channels contribute to the LSP annihilation, the impact of the
shifts in couplings could be amplified by interference effects,
leading to a significant impact on the effective annihilation
cross-section. Although we do find interference effects, 
they have little influence on the WMAP-allowed bands shown in
Fig.~\ref{fig:mh500}. In fact, contrary to what was originally
reported in \cite{Nihei:2005va}, the interference effects between
the $s$-channel Higgs and $t$-channel chargino exchange diagrams
for $\nt_1\nt_1\to W^+W^-$ are destructive, so that an enhancement
of the cross section in one channel is largely cancelled by the
other channel. An example for such an interference between the
$t$-channel $\chpm_1$ and $s$-channel $h_1$ exchange diagrams is
shown in Fig.~\ref{fig:sigvWW}.

\subsection{Annihilation through Higgs}

In the Higgs sector, nonzero phases, in particular $\phi_t$, can
induce scalar-pseudoscalar mixing as well as important changes
in the masses. One can therefore expect large differences between
the real and complex MSSM in the Higgs-funnel region.

At vanishing relative velocity, $v\to 0$,
neutralino annnihilation through $s$-channel scalar exchange
is $p$-wave suppressed; the annihilation proceeds strictly through
pseudoscalar exchange. Nevertheless when performing the thermal
averaging, the scalar exchange cannot be neglected altogether. In
the MSSM with real parameters it can amount to ${\cal O}(10\%)$ of
the total contribution. In the presence of phases, all the neutral
Higgs bosons can acquire a pseudoscalar component (that is
$g^P_{h_i{\tilde \chi^0_1}{\tilde \chi^0_1}} \neq 0$) and hence
significantly contribute to neutralino annihilation even at small $v$.
There is a kind of sum rule that relates the couplings squared of the
Higgses to neutralinos.
Therefore, for the two heavy eigenstates which are in general
close in mass, we do not expect  a large effect on the resulting
relic density from Higgs mixing alone. A noteworthy exception
occurs when, for kinematical reason, only one of the
resonances is accessible to neutralino annihilation.
That is for example the case when $m_{h_2}<2\mnt{1}\simeq m_{h_3}$.

For the analysis of the Higgs funnel, we choose
\begin{equation}
  M_1=150{\rm\;GeV},~\tan\beta=5,~
  M_S=500{\rm\;GeV},~ A_t=1200{\rm\;GeV}.
\label{eq:funnelpar}
\end{equation}
We consider the cases of small and large Higgsino mass parameter,
$\mu=500$~GeV and $\mu=1$--2~TeV, leading to small and large
mixing in the Higgs sector respectively for $\phi_t\neq 0$. In
both cases the LSP is dominantly bino. As mentioned above,
allowing for nonzero phases not only affects the neutralino and
Higgs couplings but also their physical masses. Since the relic
density is very sensitive to the mass difference $\Delta
m_{{\tilde \chi^0_1} h_i}= m_{h_i}-2{m_{\tilde \chi^0_{1}}}$
~\cite{Allanach:2004xn, Belanger:2005jk}, it is important to
disentangle the phase effects in kinematics and in couplings. As
we will see, a large part of the huge phase effects reported in
Ref.~\cite{Gondolo:1999gu} can actually be attributed to a change
in $\Delta m_{{\tilde \chi^0_1} h_i}$.

\subsubsection{Small Higgs mixing}

We fix $\mu=500{\rm\;GeV}$ so that there is small Higgs mixing.
Moreover, we set $\phi_\mu=0$ to avoid the eEDM constraint,
and discuss the dependence on $\phi_1$ and $\phi_t$.

For real parameters and $m_{H^+}=340{\rm\;GeV}$, we have
$\mnt{1}=147$~GeV, $m_{h_2}=331.5$~GeV, $m_{h_3}=332.3$~GeV, and
$\Omega h^2=0.11$. In this case, $h_2$ is the pseudoscalar. The
LSP annihilation channels are then characterized by the $h_2$
branching fractions, giving predominantly fermion pairs,
$b\bar{b}$ (78\%) and $\tau\bar\tau$ (10\%), with a small
contribution of $Zh_1$ (7\%). When we vary the phases of $A_t$
and/or $M_1$, we observe large shifts in the relic density.

\begin{figure}
  \centerline{\epsfig{file=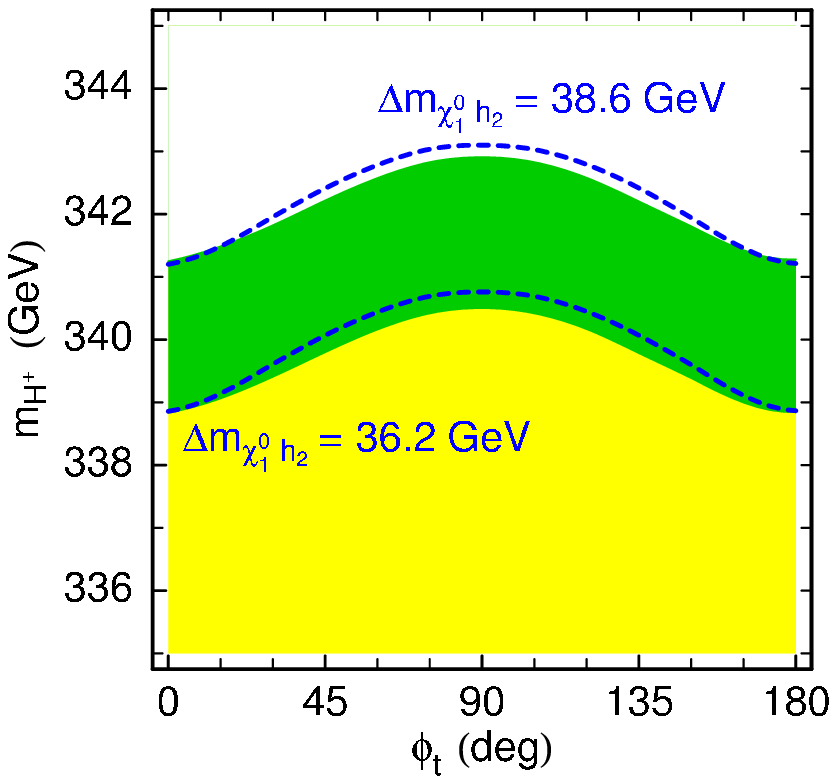, width=7cm}\hspace{6mm}
              \epsfig{file=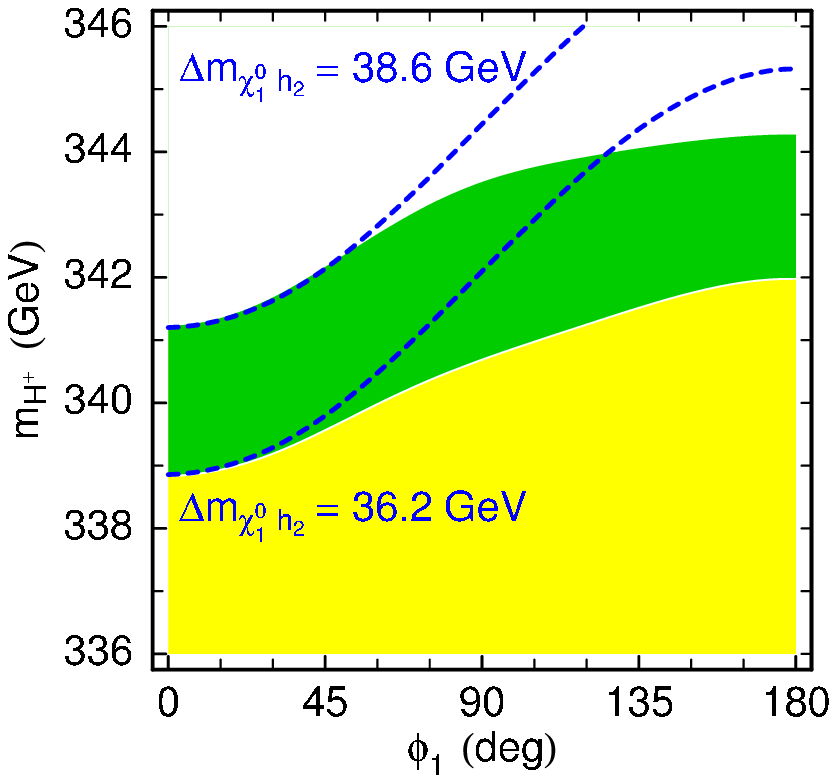, width=7cm}}
  \ablabels{3}{81}{70}
  \caption{The $2\sigma$ WMAP bands (green/dark grey) in the
  $m_{H^+}$--$\,\phi_t$ and $m_{H^+}$--$\,\phi_1$ planes for
  the parameters of Eq.~(\ref{eq:funnelpar}) and $\mu=500$~GeV;
  $\phi_1=0$ in (a) and $\phi_t=0$ in (b).
  Contours of constant mass differences
  $\Delta m_{\nt_1h_2}=m_{h_2}-2\mnt{1}$ are also displayed.
  In the yellow  (light grey) region, $\Omega h^2$ is below the WMAP range.}
\label{fig:at}
\end{figure}

First consider varying the phase $\phi_t$, which affects
the Higgs masses and mixings through loop effects. In this
scenario with relatively small $\mu$,  the scalar-pseudoscalar
mixing never exceeds 8\%.
In Fig.~\ref{fig:at}a we plot the band that is allowed by WMAP
in the $m_{H^+}$--$\phi_t$ plane. One can see that the lower and
upper WMAP bounds correspond to the contours of
$\Delta m_{{\tilde \chi^0_1} h_2}=36.2$ and $38.6$~GeV respectively,
with only $4\%$ deviation.
So the main effect of $\phi_t$ can be explained by shifts in the
physical masses and position of the resonance.

\begin{figure}
  \centerline{\epsfig{file=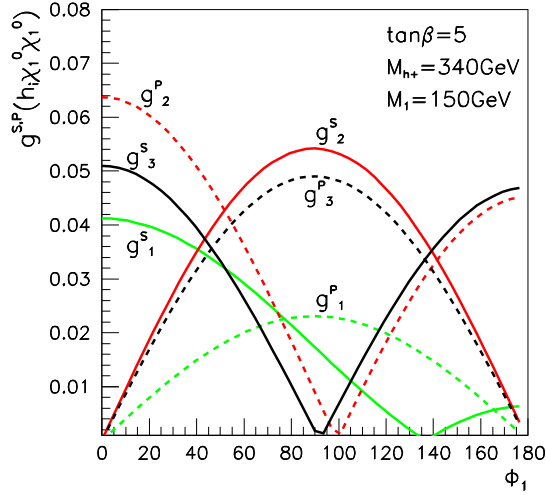, width=8.cm}}
  \caption{The scalar (full lines) and pseudoscalar (dash lines)
  neutralino Higgs couplings, $g^{S,P}_{h_i\nt_1\nt_1}$,  as a function of $\phi_1$
  for $m_{H^+}=340$~GeV, and the other parameters
  as in Fig.~\ref{fig:at}b.}
\label{fig:gsgp}
\end{figure}

\begin{figure}
  \centerline{\epsfig{file=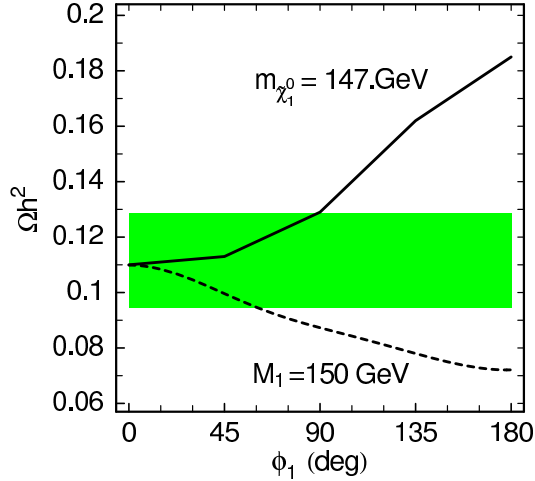, width=7cm}}
  \caption{$\Omega h^2$ as a function of $\phi_1$ for
  $m_{H^+}=340$~GeV and the value of $M_1$ adjusted so that
  $\Delta m_{{\tilde \chi^0_1} h_2}$ stays constant (full line).
  The other parameters as in Fig.~\ref{fig:at}b ($m_{h_2}=332.5$~GeV)
  For comparison, the variation of $\Omega h^2$ for fixed
  $M_1=150$~GeV is shown as a dashed line.
  The green (grey) band corresponds to the $2\sigma$ WMAP range.}
\label{fig:omega}
\end{figure}

We next  vary the phase $\phi_1$, keeping $\phi_t=0$. This changes
the neutralino masses and mixing, and hence also the
neutralino--Higgs couplings, Eq.~(\ref{eq:gs}). For
$m_{H^+}=340{\rm\;GeV}$, when increasing $\phi_1$, the relic
density drops, see Fig.~\ref{fig:at}b. This is because the mass of
the neutralino increases slowly, resulting in a smaller $\Delta
m_{{\tilde \chi^0_1} h_2}$. Adjusting $\mnt{1}$ or $m_{h_2}$ (by
changing $M_1$ or $m_{H^+}$) such that the mass difference stays
constant, we find rather that the relic density increases with
$\phi_1$. This can be readily understood from the phase dependence
of the couplings of $h_{2,3}$ to the LSP, shown in
Fig.~\ref{fig:gsgp}. For $\phi_1=0$, the coupling of $h_2$ is
predominantly pseudoscalar  and $h_3$ almost purely scalar while
for $\phi_1=90^\circ$, it is $h_3$ that has a large pseudoscalar
coupling. Therefore for $\phi_1=0$, $h_2$ exchange dominates with
a large cross section while for $\phi_1=90^\circ$ one gets about
equal contributions from $h_2$ and $h_3$, although with a smaller
overall cross section. When increasing $\phi_1$ further (up to
$180^\circ$), $h_2$ exchange again dominates, however with a
coupling to neutralinos smaller by 30\% than for $\phi_1=0$. Thus
one needs a smaller mass splitting $\Delta m_{{\tilde \chi^0_1}
h_2}$ for $\Omega h^2$ to fall within the WMAP range, see
Fig.~\ref{fig:at}b.  Moreover, for large $\phi_1$ there is also a
sizeable contribution from ${\tilde \chi^0_1} {\tilde
\chi^0_1}\rightarrow h_1 h_1$ with a constructive  interference
between $s$-channel $h_3$ and $t$-channel neutralino exchange. In
Fig.~\ref{fig:omega} we show the variation of $\Omega h^2$ with
$\phi_1$ while keeping $\Delta m_{{\tilde \chi^0_1} h_2}$ fixed.
The maximum deviation which comes purely from modifications in the
couplings can reach 70\%.

\subsubsection{Large Higgs mixing}

We next discuss the case of large mixing in the Higgs sector,
which is achieved for large values of $\mu=1$--2~TeV.
For this purpose we concentrate on the $\phi_t$ dependence.
It has to be noted that here, for large $\phi_t\approx 90^\circ$,
rather light $h^0_i$ and large $\mu A_t$, the EDM constraint
is not satisfied by simply setting $\phi_\mu=0$. One has to
either allow for a small $\phi_\mu\sim {\cal O}(1^\circ)$,
or appeal to cancellations due to light sleptons with
masses of few $\times 100$~GeV. We choose the latter solution,
imposing $\phi_\mu=0$. Adjusting the selectron parameters such that
$d_e<2.2\times 10^{-27} {\rm e\,cm}$ has only a ${\cal O}(1\%)$ effect
on the relic density in our examples.

\begin{figure}\setlength{\unitlength}{1mm}
  \centerline{\epsfig{file=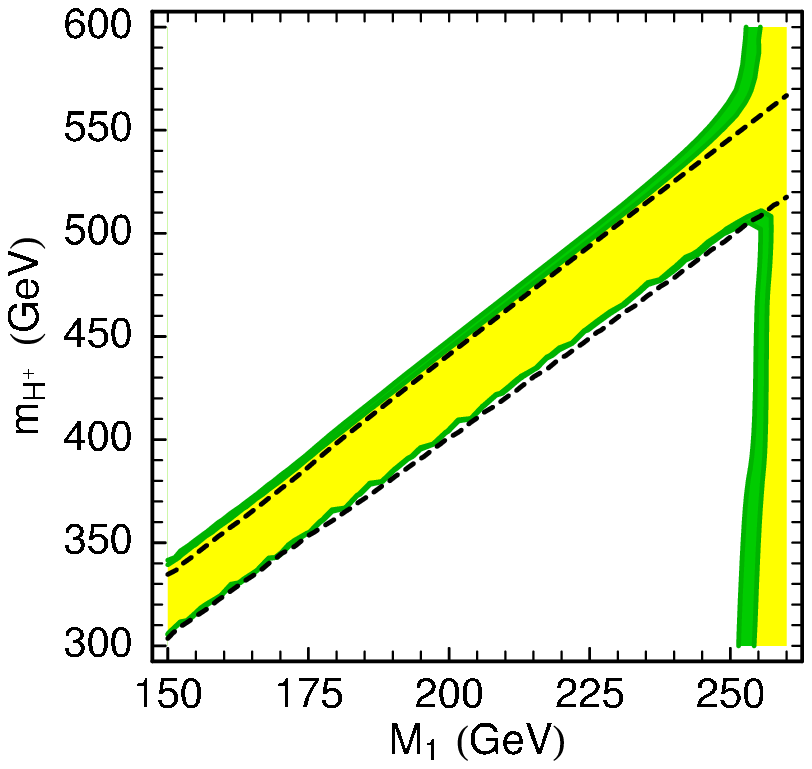, width=7cm}\hspace{6mm}
              \epsfig{file=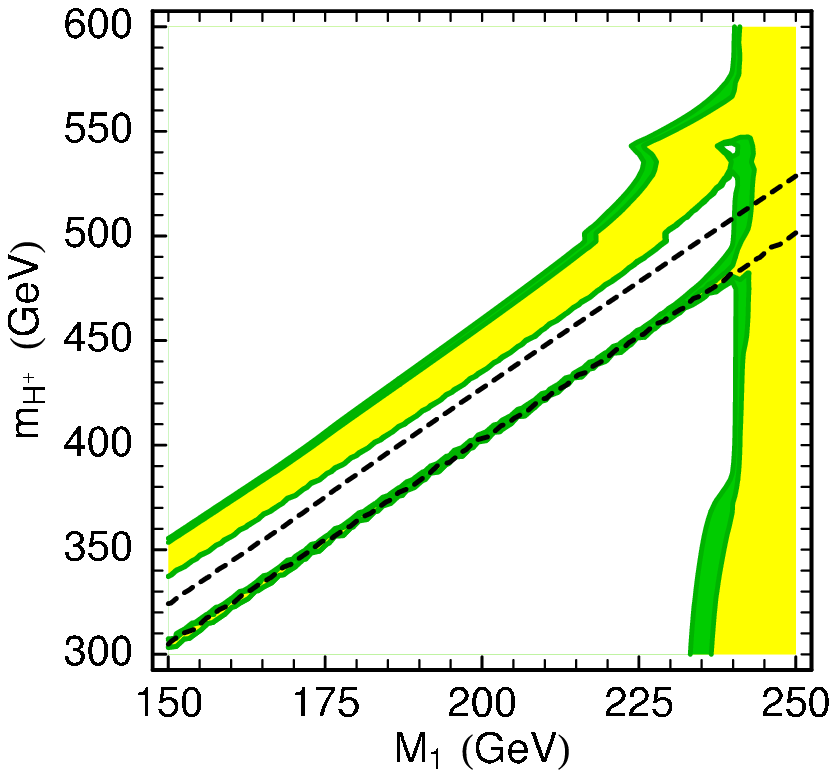, width=7cm}}
  \ablabels{3}{81}{70}
  \caption{The WMAP-allowed bands in the $m_{H^+}$--$M_1$
  plane for (a)~$\mu=1$~TeV and (b)~$\mu=2$~TeV with
  $\phi_t=90^\circ$, $\phi_1=0$ and the other parameters given by
  Eq.~(\ref{eq:funnelpar}).
  In the narrow green (dark grey) bands $0.0945\le\Omega h^2\le 0.1287$,
  while in the yellow (light grey) regions $\Omega h^2<0.0945$.
  The positions of the WMAP bands for $\phi_t=0$ are shown
  as dashed lines.}
\label{fig:omcp_funnel_high}
\end{figure}

In  Figure~\ref{fig:omcp_funnel_high} we show the
WMAP-allowed regions in the $m_{H^+}$--$M_1$ plane for this choice
of parameters and maximal phase of $A_t$ ($\phi_t=90^\circ$). The
regions for which $\Omega h^2$ falls within the WMAP band
are shown in green (dark grey), and
those for which $\Omega h^2$ is too low in yellow (light grey).
In addition, the
positions of the WMAP-allowed strips for $\phi_t=0$ are shown as
dashed lines. In the CP-conserving case, $h_3$ is a pure
pseudoscalar and $h_2$
a pure scalar, while for $\phi_t=90^\circ$ it is just the opposite
and $h_2$ is dominantly pseudoscalar. 

\begin{figure}
  \centerline{\psfig{file=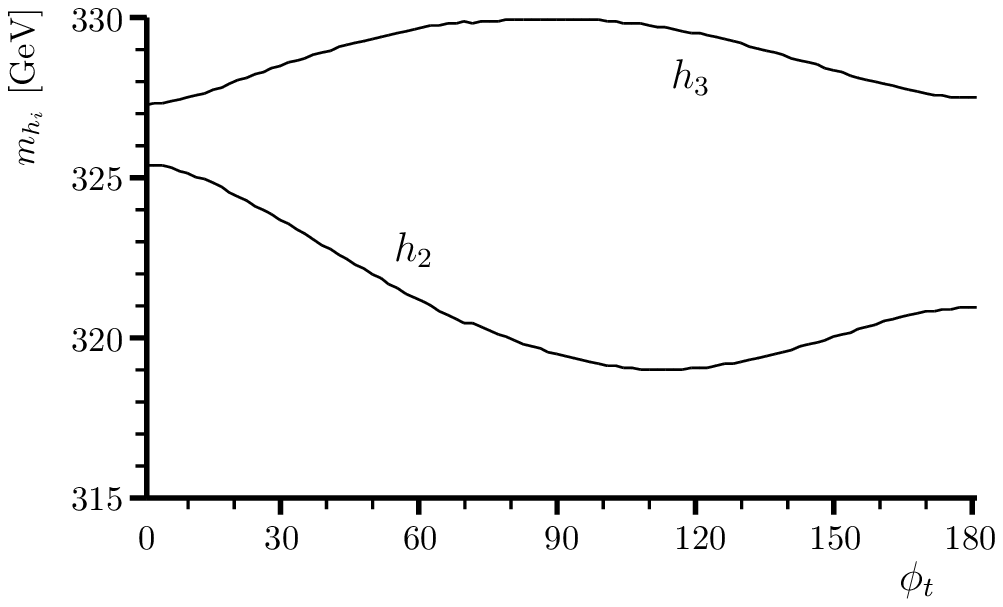, height=4.6cm}\hspace{6mm}
              \psfig{file=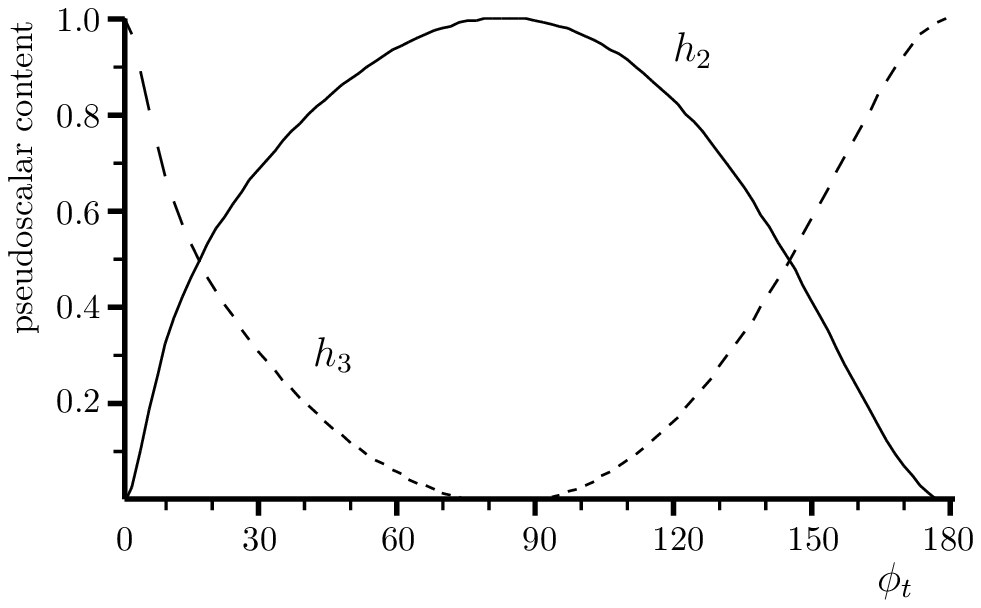, height=4.6cm}}
  \caption{Masses and pseudoscalar content of $h_2$ and $h_3$ as function
     of $\phi_t$ for $m_{H^+}=335$~GeV, $\mu=1$~TeV, and the other
     parameters given by Eq.~(\ref{eq:funnelpar}).
     The light Higgs has a mass of $m_{h_1}\simeq 117$~GeV and
     a pseudoscalar content of $\leq 10^{-4}$.}
\label{fig:ps_contents}
\end{figure}

For $\mu=1$~TeV, Fig.~\ref{fig:omcp_funnel_high}a, the mass
splitting between $h_{2,3}$ is about 10 GeV for $\phi_t=90^\circ$,
as compared to about 2 GeV for $\phi_t=0$. Masses and pseudoscalar
contents, $H_{3i}^2$, of $h_{2,3}$ are depicted in
Fig.~\ref{fig:ps_contents} as function of $\phi_t$.
Here note that it is $h_2$, i.e.\ the state which
changes from scalar to pseudoscalar with increasing $\phi_t$,
which shows the more pronounced change in mass; the crossovers of 50\%
scalar-pseudoscalar mixing of  $h_{2,3}$ occur at $\phi_t\sim
15^\circ$ and $145^\circ$. 
For $M_1$ values up to 250~GeV, we therefore find in both the 
CP-conserving and the CP-violating case two narrow bands where 
$0.094<\Omega h^2<0.129$.
For $\phi_t=0$ (and also for $\phi_t=180^\circ$) both these bands
are mainly due to pseudoscalar $h_3$ exchange, with one band just
below and the other one above the pseudoscalar resonance. For
$\phi_t=90^\circ$ the situation is different: in the lower
WMAP-allowed band the LSP annihilates through the scalar $h_3$,
with the pseudoscalar $h_2$ not accessible because
$m_{h_2}<2m_{{\tilde \chi^0_1}}\simeq m_{h_3}$, while in the upper
band both $h_2$ and $h_3$ contribute (with $h_2$ exchange of
course dominating). In between the two WMAP-allowed green bands
one is too close to the pseudoscalar resonance and $\Omega h^2$
falls below the WMAP bound; this holds for both $\phi_t=0$ and
$\phi_t=90^\circ$. The positions of the WMAP-allowed bands for
$\phi_t=0$ and $\phi_t=90^\circ$ are not very different from each
other. Still the difference in the relic density between
$\phi_t=0$ and $\phi_t=90^\circ$ is typically a factor of a few in
the WMAP-bands, and can reach orders of magnitudes at a pole. For
$M_1\gsim 250$~GeV and $\phi_t=90^\circ$, one enters the region of
coannihilation with stops, leading to a vertical WMAP-allowed
band. For $\phi_t=0$, the $\tilde t_1$ is 55~GeV heavier, so the
stop coannihilation occurs only at $M_1\sim 305$~GeV (for
$\phi_t=180^\circ$ on the other hand, $m_{\tilde t_1}\simeq
230$~GeV and coannihilation already sets in at $M_1\sim 200$~GeV).

For $\mu=2$~TeV, Fig.~\ref{fig:omcp_funnel_high}b, there is an
even stronger CP-mixing of $h_{2,3}$ and the mass splitting
between the two states becomes $\sim 45$~GeV for
$\phi_t=90^\circ$. The pseudoscalar contents are similar to those
in Fig.~\ref{fig:ps_contents} (r.h.s.\ plot) with the 50\% cross-over at
$\phi_t\sim 20^\circ$. Moreover, because the LSP has less Higgsino
admixture, one has to be closer to resonance to obtain the right
relic density. As a result, the scalar and pseudoscalar funnels
become separated by a region where $\Omega h^2$ is too
large~\cite{Choi:2006hi}.
In fact both the $h_2$ and $h_3$ exchange each lead to two WMAP-allowed
bands, one above and one below the respective resonance. For the
$h_3$ (scalar) exchange, however, these two regions are so close
to each other that they appear as one line in
Fig.~\ref{fig:omcp_funnel_high}b. This is in sharp contrast to the
CP-conserving case, $\phi_t=0$, where the scalar and pseudoscalar
states are close in mass, hence leading to only two WMAP-allowed
bands. These are again shown as dashed lines in
Fig.~\ref{fig:omcp_funnel_high}b and origin dominantly from the
pseudoscalar resonance, the scalar resonance being `hidden'
within.

\begin{figure}
  \centerline{\epsfig{file=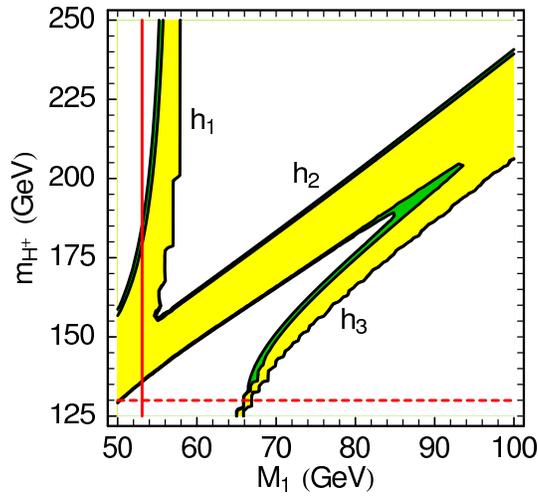, width=7cm}}
  \caption{Same as Fig.~\ref{fig:omcp_funnel_high}a but for light
  neutralinos and Higgs bosons. The vertical red line indicates
  the LEP bound on the $\ch_1$ mass, the horizontal dashed line that
  on $m_{h_1}$.}
\label{fig:mhm1low}
\end{figure}

We also display in Fig.~\ref{fig:mhm1low} the WMAP-allowed bands
for $\mu=1$~TeV focusing on the region of small neutralino and
Higgs masses. Here we see clearly three specific Higgs
annihilation funnels, each delimited by two narrow bands where
$\Omega h^2$ is within the WMAP bound. The band corresponding to
$h_1$ exchange is also found in the CP-conserving case. However,
in the CP-conserving case the LEP limit \cite{Barate:2003sz} on
the Higgs mass, $m_{h^0}>114$~GeV, requires $m_{H^+}\gsim
210$~GeV, while in the CP-violating case the limit is much lower,
about $m_{h_1}\gsim 85$~GeV~\cite{Bechtle:2006iw} at $\tb=5$ and
$m_t=175$~GeV, corresponding to $m_{H^+}\gsim 130$~GeV. For
$m_{H^+}\lsim 190$~GeV, the bands corresponding to $h_2$ and $h_3$
exchange are clearly separated because here the mass splitting
between the two states is large, about 20--36~GeV. This is to be
contrasted with the real case, where the $H/A$ mass splitting is
much smaller and annihilation through the pseudoscalar always
dominates.

\begin{figure}\setlength{\unitlength}{1mm}
  \centerline{\epsfig{file=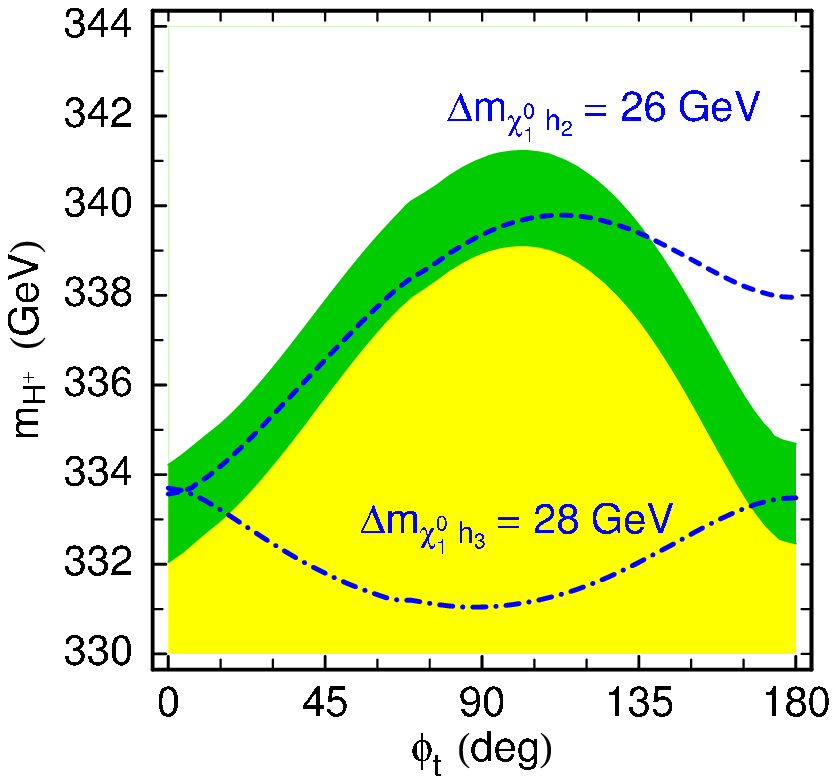, width=7cm}\qquad
              \epsfig{file=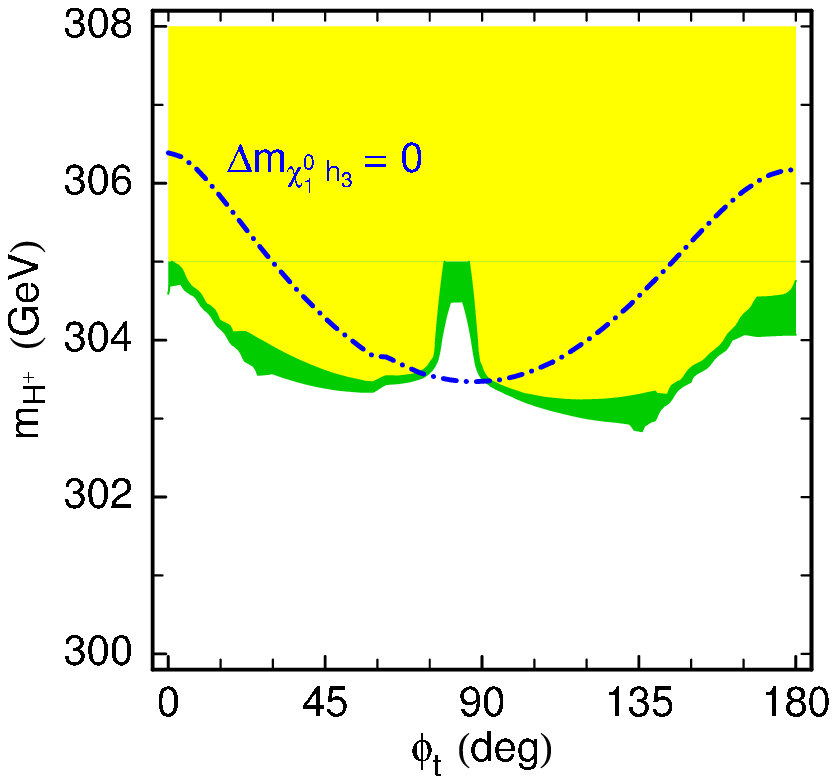, width=7cm}}
  \ablabels{3}{81}{70}
  \caption{The $2\sigma$ WMAP bands (green/dark grey) in the
  $m_{H^+}$--$\phi_t$ plane for $\mu=1$~TeV and
  $m_{H^+}\sim 330$--340~GeV in (a) and $m_{H^+}\sim 305$~GeV
  in (b). Contours of constant mass differences
  $\Delta m_{{\tilde \chi^0_1} h_i}$ are also
  displayed.  In the yellow (light grey) region, $\Omega h^2$ is
  below the WMAP range.}
\label{fig:scenario2}
\end{figure}

Let us now examine the dependence on $\phi_t$ in more details.
For this we fix $\mu=1$~TeV. 
For vanishing phases, agreement with WMAP is found for
$m_{H^+}\simeq 332$--334~GeV. This value is lower than in the
scenario with small Higgs mixing because the Higgsino fraction of
the LSP is smaller, so one needs to be closer to the Higgs
resonance. For $\phi_t\neq 0$ we have a large scalar-pseudoscalar
mixing and hence a stronger dependence of $\Omega h^2$ on
$\phi_t$. For $\phi_t=0$, $h_3$ is the pseudoscalar and gives the
dominant contribution to neutralino annihilation while for
$\phi_t=90^\circ$ $h_2$ is the pseudoscalar, hence giving the
dominant contribution. Consequently in Fig.~\ref{fig:scenario2},
agreement with WMAP is reached for $\Delta m_{{\tilde \chi^0_1}
h_i}\sim 25{\rm\;GeV}$ with $h_i=h_3$ at $\phi_t=0$ and
$180^\circ$, and $h_i=h_2$ at $\phi_t=90^\circ$.

When twice the LSP mass is very near the heaviest Higgs resonance,
one finds another region where the relic density falls within the
WMAP range. This is shown in Fig.~\ref{fig:scenario2}b
(corresponding to the phase dependence of the lower WMAP-allowed
band in Fig.~\ref{fig:omcp_funnel_high}a).
In the real case one needs
$m_{H^+}=305$~GeV, giving a mass difference $\Delta m_{{\tilde
\chi^0_1} h_3}=-1.5{\rm\;GeV}$. Note that annihilation is
efficient enough even though one catches only the tail of the
pseudoscalar resonance. For the same charged Higgs mass, the mass
of $h_3$ increases when one increases $\phi_t$, so that neutralino
annihilation becomes more efficient despite the fact that $h_3$
becomes scalar-like and $g^P_{{\tilde \chi^0_1}{\tilde \chi^0_1}
h_3}$ decreases. When $\phi_t\sim75^\circ$--$90^\circ$, the coupling
$g^P_{{\tilde \chi^0_1}{\tilde \chi^0_1} h_3}$ becomes very small
and one needs $\Delta m_{{\tilde \chi^0_1} h_3}=0$--$1.5$~GeV
to achieve agreement with WMAP, see Fig.~\ref{fig:scenario2}b.
Here we are in the special case where $m_{h_2} < 2m_{{\tilde
\chi^0_1}} \simeq m_{h_3}$, so that only $h_3$ contributes
significantly to the relic density.

We can isolate the phase dependence of $\Omega h^2$ due to the
scalar-pseudoscalar mixing by keeping the distance from the $h_3$
pole constant. For constant values of $\Delta
m_{{\tilde \chi^0_1} h_3}=-1.5{\rm\;GeV}$ we get an increase in
$\Omega h^2$ relative to the $\phi_t=0$ case by almost an order of
magnitude. This is however far less than the huge shifts of several 
orders of magnitude found for fixed values of $m_{H^+}$ when a
Higgs pole is passed.

\subsection{Bino-like LSP and light sfermions}

In the CP-conserving MSSM, light sfermions can significantly contribute
to reducing the relic density to a value which is in agreement with WMAP, 
in particular in the case of a bino-like LSP.
The relevant processes are $\nt_1\nt_1\to f\bar f$ via $t$-channel
sfermion exchange, as well as coannihilation with sfermions that
are close in mass to the LSP.
In the CPV-MSSM with large phases, the sfermions of the first and
second generation need to be heavy to avoid the EDM constraints.
The third generation is however much less constrained and can be light.
We therefore consider in this section the cases of light staus
and light stops.

\subsubsection{Light staus}

We choose a scenario where the LSP is mostly bino and fix
$\tan\beta=10$, $\mu=600$~GeV, $m_{H^+}=M_S=A_{f,t}=1$~TeV.
Moreover, we take $M_{\tilde R_3}=220$~GeV and $M_{\tilde L_3}=240$~GeV,
so that staus are relatively light, $m_{\ti\tau_1}=212.1$~GeV,
and can contribute to the neutralino annihilation.
Note that for this choice of parameters there is a large mixing
in the stau sector, driven by $\mu\tan\beta$.
The eEDM constraint is avoided by setting $\phi_\mu=\phi_t=\phi_l=0$.
To obtain a relic density in agreement with WMAP one has to rely
on stau (co)annihilation channels. For this aim, the mass difference
$\Delta m_{\nt_1\ti\tau_1}=m_{\ti\tau_1}-\mnt{1}$ must be small,
in our example about 4--6~GeV.
For $M_1=210$~GeV and vanishing phases, we obtain $\mnt{1}=208$~GeV,
$\Delta m_{\nt_1\ti\tau_1}=4.1$~GeV and $\Omega h^2=0.102$.
The main channels in this case are
$\nt_1\ti\tau_1\to\gamma\tau$ (32\%),
$\nt_1\ti\tau_1\to Z\tau$ (10\%),
$\ti\tau_1\ti\tau_1\to\tau\tau$ (26\%) and
$\ti\tau_1\ti\tau_1^*\to\gamma\gamma$ (12\%).

\begin{figure}
  \centerline{\epsfig{file=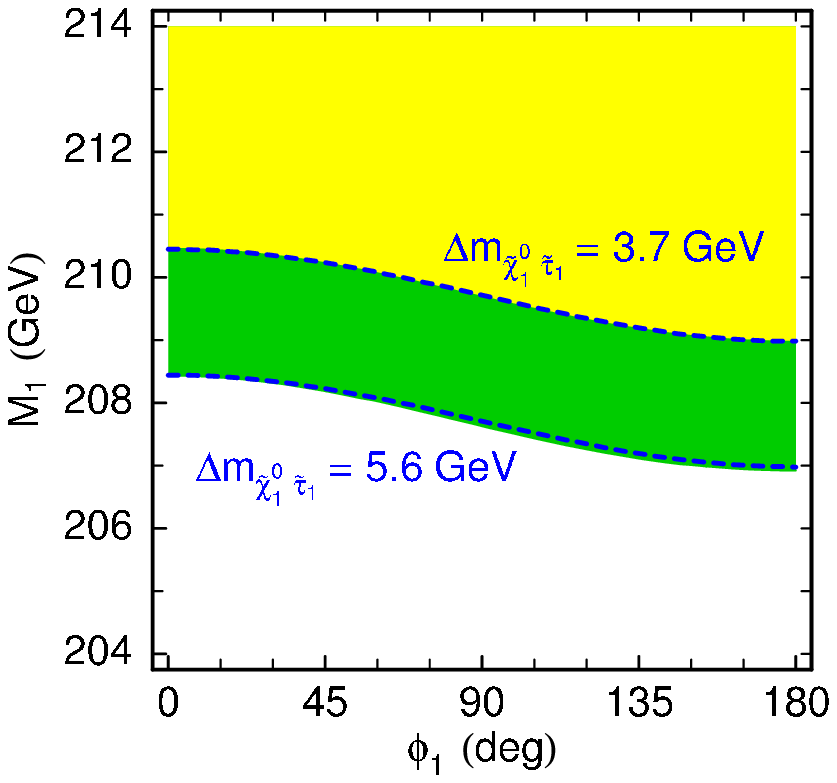, width=7cm}\qquad
              \epsfig{file=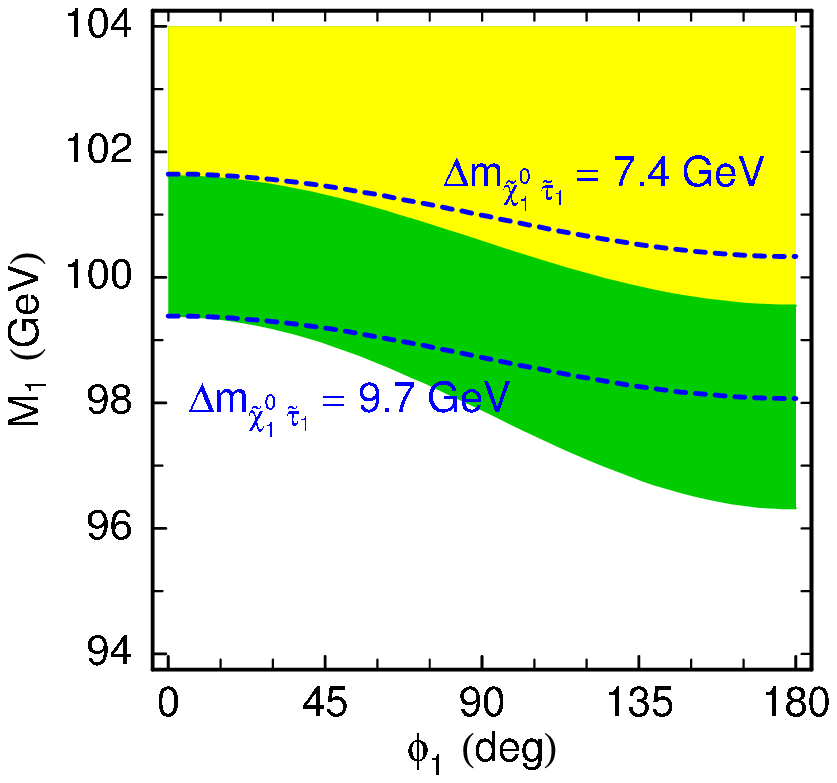, width=7cm}}
  \ablabels{2}{80}{70}
  \caption{The $2\sigma$ WMAP bands (green/dark grey) in the
  $M_1$--$\phi_1$ plane for $m_{\tilde\tau_1}\simeq 212$~GeV in (a)
  and $m_{\tilde\tau_1}\simeq 108$~GeV in (b). The other parameters
  are given in the text.
  In the yellow (light grey) region, $\Omega h^2$ is below the WMAP bound.
  Contours of constant mass difference $\Delta m_{\nt_1\tilde\tau_1}$
  are shown as dashed lines.}
\label{fig:stau}
\end{figure}

It is well known that the mass difference is the key parameter in
case of coannihilations. We therefore expect large shifts in
$\Omega h^2$ for nonzero phases, resulting from small changes in
the masses. Figure~\ref{fig:stau}a shows the WMAP-allowed region
in the $M_1$--$\phi_1$ plane for the scenario given above. As can
be seen, the WMAP band matches almost perfectly with the contours
of constant mass difference, $\Delta m_{\nt_1\tilde\tau_1}=3.7$
and $5.6$~GeV. When we adjust the parameters of the stau sector to keep
a constant mass difference while varying $\phi_1$, the relic density 
stays constant within few percent, $\delta\Omega/\Omega\lsim 5\%$.
An analogous behaviour is found for nonzero $\phi_l$. 

For very light neutralinos and sleptons, annihilation into lepton
pairs through $t$-channel slepton exchange can be efficient enough
to achieve $\Omega h^2\sim 0.1$. In the mSUGRA model, this is
often called the `bulk' region. Owing to the LEP limit of
$m_{\ti\l}\gsim 90$--100~GeV \cite{LEPSUSY} (depending on the
slepton flavour and chirality/mixing and on $m_{\ti l}-\mnt{1}$),
and because the $t$-channel slepton contribution scales as
$\mnt{1}^2/m_{\ti l}^4$, the `bulk' is squeezed into a small
region of slepton masses of about 100~GeV. To investigate this
case in the CPV-MSSM, we lower the stau parameters to $M_{\tilde
R_3}=135$~GeV, $M_{\tilde L_3}=150$~GeV and $A_\tau=100$~GeV,
$\phi_l=0$. This gives $m_{\tilde\tau_1}=107.8$~GeV,
$m_{\tilde\tau_2}=182.1$~GeV and $m_{\tilde\nu_\tau}=135.7$~GeV.
The WMAP band in the $M_1$--$\phi$ plane for this scenario is
shown in Fig.~\ref{fig:stau}b. For $\mnt{1}\lsim 100$~GeV, that is
up to $M_1\sim 100$~GeV, $\nt_1\nt_1\to \tau^+\tau^-$ is the
dominant process. For heavier neutralinos, $\nt_1\ti\tau_1$
coannihilation dominates. Agreement with WMAP is achieved for
larger $\nt_1$--$\ti\tau_1$ mass differences than in the previous
example. Since coannihilations are less important, we find a
stronger phase dependence which is not completely determined by
$\Delta m_{\nt_1\ti\tau_1}$, see Fig.~\ref{fig:stau}b. When
keeping the masses constant, the maximal variation in $\Omega h^2$
due to $\phi_1$ is about 15\%. Last but not least note that
$s$-channel $Z$ and Higgs exchange is negligible in this scenario;
$t$-channel exchange of $\ti\tau_2$ however does contribute and
there is in fact a strong destructive interference between the
$\ti\tau_1$ and $\ti\tau_2$ exchange diagrams.

\begin{figure}
  \centerline{\epsfig{file=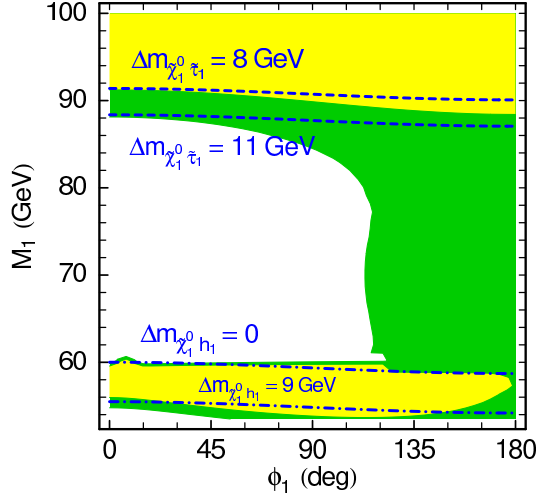, width=7cm}}
  \caption{The $2\sigma$ WMAP bands (green/dark grey) in the
  $M_1$--$\phi_1$ plane for $m_{\tilde\tau_1}\simeq 98$~GeV.
  The other parameters are given in the text.
  In the yellow (light grey) region, $\Omega h^2$ is below the WMAP bound.
  Also shown are contours of constant mass differences
  $\Delta m_{\nt_1\tilde\tau_1}=m_{\ti\tau_1}-\mnt{1}$ (dashed lines)
  and $\Delta m_{\nt_1h_1}=m_{h_1}-2\mnt{1}$ (dash-dotted lines).
  }
\label{fig:stau_low}
\end{figure}

Let us lower the stau masses parameters even further, close to the
experimental limit while keeping $\phi_\tau=0$. For $M_{\tilde
R_3}=130$~GeV, $M_{\tilde L_3}=140$~GeV, we get
$m_{\tilde\tau_1}=98.2$~GeV, $m_{\tilde\tau_2}=175.9$~GeV and
$m_{\tilde\nu_\tau}=124.6$~GeV. The WMAP-allowed regions in the
$M_1$--$\phi_1$ plane for this case are shown in
Fig.~\ref{fig:stau_low}. The almost horizontal bands of
annihilation through the light Higgs ($m_{h_1}=118$~GeV) for very
light $\nt_1$, as well as the stau coannihilation region at
$M_1\gsim 90$~GeV are clearly visible. The peculiar feature is
that for large phase, $\phi_1\gsim 110^\circ$, there appears a new
region, connecting the light Higgs funnel and the stau
coannihilation strip, where $\Omega h^2$ is within the WMAP bound.
In this region, $\nt_1\nt_1\to\tau\tau$ completely dominates, and
the phase dependence originates from the $\nt_1\stau_{1,2}\tau$
couplings. Again, there is an important interference between the
$\stau_1$ and $\stau_2$ exchange diagrams.

\begin{figure}
  \centerline{\psfig{file=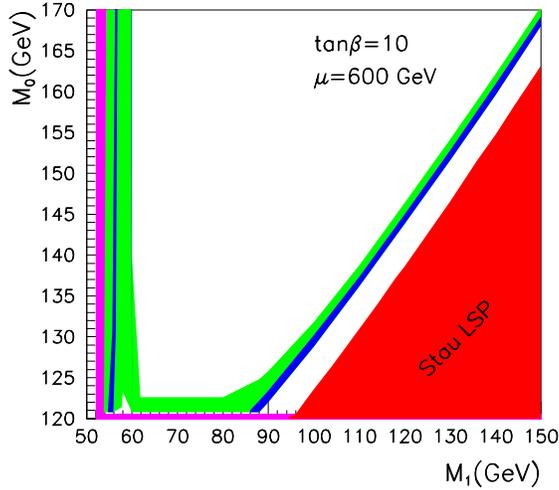, width=8cm}}
  \caption{The $2\sigma$ WMAP bands  in the
  $M_0$--$M_1$ plane, with $M_0=M_{\ti R_3}=0.9M_{\ti L_3}$;
  the blue (black) bands are for vanishing phases, the green
  (medium grey) one for arbirary phases, $\phi_1,\phi_t=\phi_l$.
  The pink (dark grey)
  region is excluded by the requirements $\mch{1}>103.5$~GeV, $m_{\stau_1}>95$~GeV.
  In the lower right-hand corner (red/dark grey) the $\stau_1$ is the
  LSP.
  }
\label{fig:stau_m0m1}
\end{figure}

That this new `stau bulk' region is generic can be seen in
Fig.~\ref{fig:stau_m0m1}. Here we plot the WMAP-allowed bands in
the $M_0$--$M_1$ plane, with $M_0\equiv M_{\tilde
R_3}=0.9M_{\tilde L_3}$.  For vanishing phases, the light Higgs
funnel and $\stau$ coannihilation regions appear as narrow
disconnected strips. For arbitrary phases $\phi_1$,
$\phi_\mu$, $\phi_t$, $\phi_l$, these strips are much
wider; especially the light Higgs funnel becomes a band instead of 
a narrow strip. Furthermore, the Higgs funnel  
is connected to the $\stau$ coannihilation region
by the $t$-channel stau exchange region, which appears as a horizontal
band at $M_0\sim 130$~GeV.

\subsubsection{Light stops}

To discuss the case of a light stop, we fix
$M_{\tilde Q_3}=500$~GeV, $M_{\tilde U_3}=450$~GeV,
$M_{\tilde D_3}=800$~GeV, $\mu=M_S=m_{H+}=1$~TeV,
and $\tb=5$. We again fix $\phi_\mu=0$ to easily avoid the eEDM constraint.
Moreover, we choose $A_t=1$~TeV and $\phi_t=180^\circ$
to obtain a light $\st_1$, $\mst{1}=243.5$~GeV.
Since $\mu$ is large we are again in a scenario with a bino LSP.
A relic density in agreement with WMAP is found for $M_1\approx 215$~GeV
($\mnt{1}\approx 214$~GeV) in the real case. The main channels
are $\nt_1\nt_1\to t\bar t$ (21\%),
$\nt_1\st_1\to th_1$ (57\%), and $\nt_1\st_1\to gt$ (17\%).
Note that $\nt_1\st_1$ coannihilation dominates although the mass
difference is much larger than in the case of $\nt_1\ti\tau_1$
coannihilation.

The phases that can play a role here are $\phi_1$ and $\phi_t$.
As we have already observed in other coannihilation scenarios,
the relic density is extremely sensitive to the mass difference
$\Delta m_{\nt_1\st_1}=\mst{1}-\mnt{1}$.
However, in the stop coannihilation scenario we also observe some
important effects due to the phase dependence of the couplings.

\begin{figure}
  \centerline{\epsfig{file=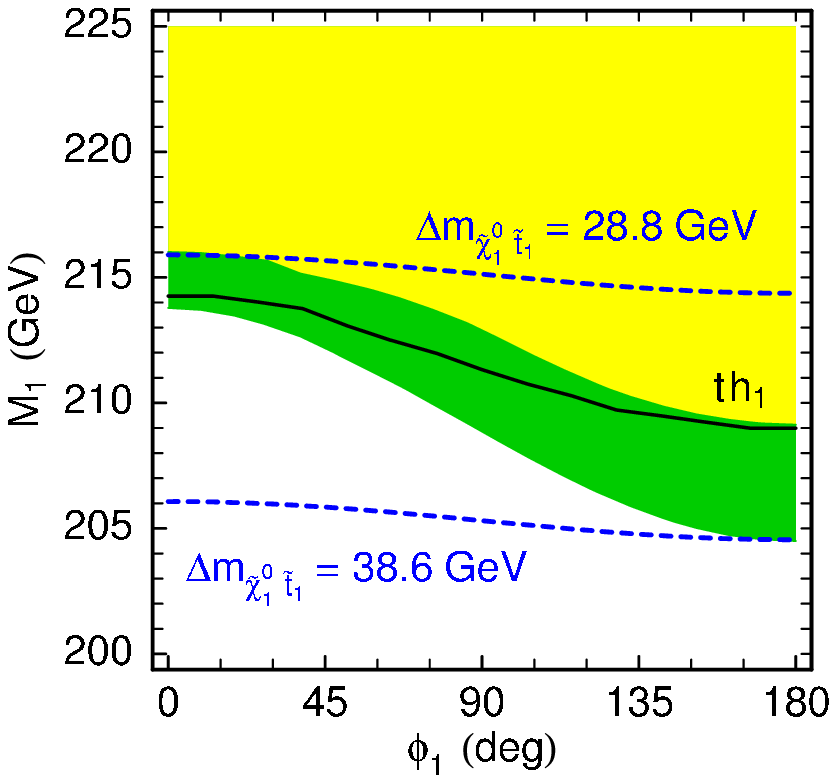, width=7cm}\qquad
              \epsfig{file=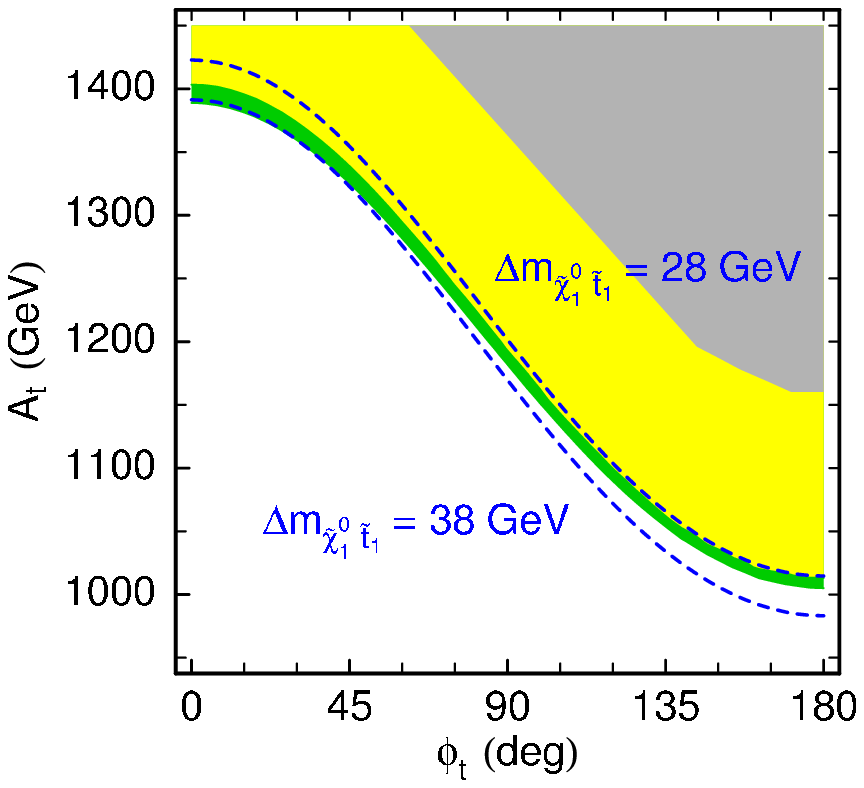, width=7cm}}
  \ablabels{2}{82}{70}
  \caption{The $2\sigma$ WMAP bands (green/dark grey) for the
  stop coannihilation scenario,
  (a) in the $M_1$--$\phi_1$ plane for $\mst{1}=243.5$~GeV
  and (b) in the $A_t$--$\phi_t$ plane for $\mnt{1}=210.8$~GeV.
  The other parameters are given in the text. In the yellow (light grey)
  regions, $\Omega h^2<0.0945$. The medium grey region in (b)
  is theoretically excluded.
  Also shown are contours of constant mass difference
  $\Delta m_{\nt_1\tilde t_1}$ (blue dashed). The full black line in (a) is
  a contour of constant $\langle\sigma v\rangle(\nt_1\st_1\to t h_1)$.}
\label{fig:stop}
\end{figure}

First we vary only $\phi_1$ and show in Fig.~\ref{fig:stop}a
the WMAP-allowed band in the $M_1$--$\phi_1$ plane.
We find the WMAP band does not match the contours of constant
mass difference $\Delta m_{\nt_1\st_1}=\mst{1}-\mnt{1}$.
A much larger mass difference is required
at $\phi_1=180^\circ$ ($\Delta m_{\nt_1\st_1}=34.2$--$38.6$~GeV) then
at $\phi_1=0$ ($\Delta m_{\nt_1\st_1}=28.8$--$31.0$~GeV).
The reason for this is an increase of both the
$\nt_1\nt_1\rightarrow t\bar{t}$ as well as the
$\nt_1\st_1\rightarrow t h_1$ cross-sections with the phase $\phi_1$.
For the coannihilation process, the phase dependence is
enhanced by a constructive interference between the
$t$-channel $\st_1$ and the $s$-channel top  exchange diagrams.

To investigate also the dependence on $\phi_t$, we now fix
$M_1=212$~GeV and $\phi_1=0$ ($\mnt{1}=210.8$~GeV)
and plot in Fig.~\ref{fig:stop}b the WMAP-allowed band in the
$A_t$--$\phi_t$ plane. The other parameters are as above.
As before, agreement with WMAP is found only for a narrow
band in which $\nt_1\st_1\rightarrow t h_1$ dominates.
Although in this band the $\st_1$ mass is constant within 10~GeV,
there is no exact match between the contours of constant
$\Omega h^2$ and $\Delta m_{\nt_1\st_1}$. Rather, at $\phi_t=0$,
$90^\circ$ and $180^\circ$,
agreement with WMAP requires $\Delta m_{\nt_1\st_1}\sim 36$~GeV,
32~GeV and 30~GeV, respectively (each about $\pm 2$~GeV).

\begin{figure}
  \centerline{\epsfig{file=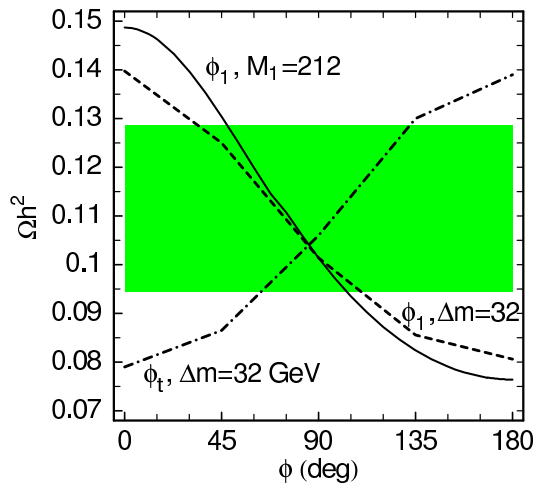, width=7cm}}
  \caption{$\Omega h^2$ in the stop coannihilation scenario as a
  function of $\phi_1$ (dashed line) and as a function of $\phi_t$
  (dash-dotted line) for constant mass difference
  $\Delta m_{\nt_1\st_1}=32$~GeV.
  For comparison, the variation of $\Omega h^2$ with $\phi_1$ for
  fixed $M_1=212$~GeV is also shown. See text for details.
}
\label{fig:stopvarphi}
\end{figure}

When keeping the mass difference constant, we can single out
the phase dependence of $\Omega h^2$ that comes solely from
changes in the couplings. This is shown in Fig.~\ref{fig:stopvarphi},
where we plot $\Omega h^2$ as a function of $\phi_1$ for $\phi_t=180^\circ$
(dashed line) and as a function of $\phi_t$ for $\phi_1=0$
(dash-dotted line), each time keeping $\Delta m_{\nt_1\st_1}=32$~GeV
constant by adjusting either $M_1$ or $A_t$.
As can be seen, in either case $\Omega h^2$ varies between 0.08 and 0.14.
For comparison, the variation of $\Omega h^2$ with $\phi_1$ for
$M_1=212$~GeV ($A_t=1$~TeV, $\phi_t=180^\circ$),
i.e.\ varying mass difference, is also shown.

\subsection{Relaxing the gaugino GUT relation: the bino/wino scenario}

We now consider relaxing the universality condition amongst the
first two gaugino mass parameters and treat $M_1$ and $M_2$ as two
independent parameters. As we want to examine specifically the role
of the wino component we fix $\mu=1$~TeV and $m_{H+}=1$~TeV, then the
Higgsino component of the LSP is small and the annihilation near
Higgs resonance not possible. Because we choose sfermions to be heavy,
the annihilation into fermion pairs is suppressed.
The LSP can hence only pair-annihilate into gauge bosons.

Choosing $M_2/M_1<2$ increases the wino component of the LSP.
Since the neutralino-chargino-$W$
coupling has also a term proportional to $N_{i2}$, one could
think that the bino-wino scenario is quite similar to the mixed
bino-Higgsino case discussed in Section~\ref{sec:Higgsino}.
However, when the parameters are set such that $f_W=|N_{i2}|^2$
becomes sizable, say around 10\%, the mass difference
$\Delta m_{\nt_1\ch_1}$ becomes small (few GeV), and the coannihilation
channels are so important that the relic density is well below the
WMAP limit, unless $M_{1,2}\sim{\cal O}(1)$~TeV.

\begin{figure}
  \centerline{\epsfig{file=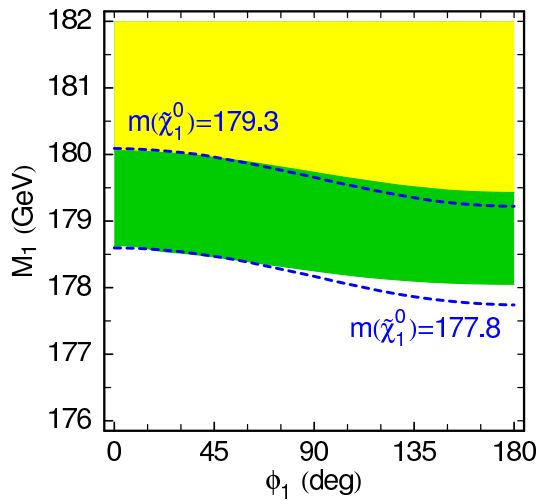, width=7cm}}
  \caption{The $2\sigma$ WMAP-allowed band (green/dark grey) in the
  $M_1$--$\phi_1$ plane for $M_2=200$~GeV, $\tb=10$,
  $\mu=M_S=m_{H+}1$~TeV, $\phi_t=\phi_\mu=0$. Superimposed are contours
  of constant $\nt_1$ masses (blue dashed lines).
  In the yellow (light grey) region, $\Omega h^2$ is
  below the WMAP range.}
\label{fig:winophi1}
\end{figure}

In fact, 
in the mass range which is interesting for collider searches, 
the LSP has to be still
overwhelmingly bino, and the relic density is completely dominated
by coannihilation channels involving $\nt_1\chpm_1$, $\nt_1\nt_2$,
$\nt_2\chpm_1$, $\nt_2\nt_2$, $\chpm_1\chpm_1$.\footnote{This
scenario was considered in Ref.~\cite{Baer:2006dz} for negative
values of $M_1$, that is $\phi_1=180^\circ$.} Final states involve
a variety of channels from gauge boson pairs to fermion pairs of
all three generations. Even when letting all phases vary, the
relic density falls within the WMAP range only for a very narrow range
of parameters; for a given value of $M_1$ the viable range of
$M_2$ varies by only 2--3~GeV. For $\tan\beta=10$, for instance,
the allowed band can roughly be parameterized as
\begin{equation}
   M_2 \simeq (2.03 \times 10^{-4} M_1^2 +1.02 M_1+9 ) \pm 2
\end{equation}
with $M_1$ given in GeV.
This is typical for scenarios that are dominated by coannihilation
in the sense that only a narrow range of mass difference between
the NLSP and the LSP is allowed. Furthermore,
considering the large number of contributing channels, the phase
dependence in each individual channel tends to be ``softened". For
example for $M_2=200$~GeV, $M_1=179$~GeV, $\tan\beta=10$, we have
$\Delta m_{\ch_1\nt_1}=21.2$~GeV and $\Omega h^2=0.121$; the dominant
channels are $\nt_2\ch_1\ra f\bar{f}'$ altogether contributing
around 40\% of the total effective annihilation cross-section. In
Fig.~\ref{fig:winophi1}, we display the WMAP allowed range in the
$M_1$--$\phi_1$ plane for $\phi_\mu=0$. The contours of constant
$\Omega h^2$ basically follow the contours of constant mass difference
$\Delta m_{\nt_1\ch_1}$, as expected when coannihilation channels
are dominant. Only for $\phi_1>90^\circ$ there is a small increase
in $\Omega h^2$ due to shifts in couplings, introducing a small
gap between the contours of constant mass difference and those of
constant $\Omega h^2$. For a given $\Delta m_{\nt_1\ch_1}$, the
maximal deviation from the case of vanishing phases reaches
$\Delta\Omega /\Omega \approx 25\%$.

\section{Summary and Conclusions}

We have performed the first complete computation of the relic density
of neutralino dark matter in the CPV-MSSM, including in a consistent
manner phases in all annihilation and coannihilation processes.
Moreover, we have presented a comprehensive study of the typical
scenarios that predict a relic density in agreement with WMAP.
Since CP phases do not only change the sparticle and Higgs mixings
but also the masses, we have taken care to disentangle effects from
kinematics and couplings.

A priori one could think that taking out the effects which come
from changes in the masses would diminish the phase dependence of
$\Omega h^2$. For processes, for which the mass difference is the
most important parameter, i.e.\ for coannihilations or
annihilation near a pole, this is indeed the case. One the other
hand, we have found several examples where the phase dependence of
masses and couplings work against each other, and taking out the
kinematic effects actually enhances the variation in $\Omega h^2$.
This happens, for instance, in the case of a mixed bino-Higgsino
LSP, where we have found effects of almost an order of magnitude
from modifications in the couplings due to nonzero CP phases. In
the case of annihilation through $s$-channel Higgs exchange, for
small scalar-pseudoscalar mixing, we have found cases where for
fixed MSSM parameters $\Omega h^2$ goes down with increasing
$\phi_1$, but when keeping the mass difference between the
pseudoscalar-like Higgs pole and 2$\mnt{1}$ constant, $\Omega h^2$
actually goes up. This effect is of the order of 50\% and
dominantly due to changes in the scalar-/pseudoscalar-type
coupling of the LSP. Much larger effects have been found for large
scalar-pseudoscalar mixing. In the peculiar case that only one
resonance is accessible to neutralino annihilation, we have found
order-of-magnitude variations in $\Omega h^2$ due to changes in
the pseudoscalar content of this resonance. Moreover, in some
cases we have found large interference effects in the most
important annihilation or coannihilation channels.
Such interferences do, however, not necessarily lead to a large 
variation in $\Omega h^2$.

We have also considered scenarios with a bino-like LSP and light sfermions
(stops and staus), and studied the phase dependence in the regions where
a)~annihilation into $f\bar f$ dominates, b)~coannihilations dominate,
as well as the intermediate region where c)~both $t$-channel exchange
and coannihilation are important.
In the stop-coannihilation region, when $\nt_1\st_1\to t h_1$ is the
dominant process, the variation in $\Omega h^2$ can be about due to
changes in the $\nt_1\st_1h_1$ couplings. There is moreover a constructive
interference between the  $t$-channel $\st_1$ and the $s$-channel top
exchange diagrams. Another peculiar feature arises for (very) light staus:
here we have found a region where for large phase $\phi_1$
the annihilation into $\tau\tau$ alone can be efficient enough to obtain
agreement with the WMAP bound. This region does not occur for zero phases.

We emphasize that even in scenarios which feature a modest phase
dependence once the kinematic effects are singled out, the
variations in $\Omega h^2$ are comparable to (and often much
larger than) the $\sim 10\%$ range in $\Omega h^2$ of the WMAP
bound.
Therefore, when aiming at a precise prediction of the neutralino relic
density from collider measurements, it is clear that one does not only
need precise sparticle spectroscopy --- one also has to precisely measure
the relevant couplings. As we have shown in this paper, this programme
certainly includes the determination of possible CP phases.
For zero phases, in a bino scenario with light sleptons,
one may be able to infer $\Omega h^2$ of the LSP with roughly the
WMAP precision ($\sim 15\%$) from LHC measurements~\cite{Nojiri:2005ph,
Arnowitt:2006jq}. At the ILC, one expects to achieve much higher precisions,
allowing for a prediction of $\Delta\Omega/\Omega$ of the level of
few percent in case of a bino LSP annihilating through
light sleptons \cite{Baltz:2006fm} and
in the stau coannihilation scenario~\cite{Martyn:2004jc,Bambade:2004tq}.
In the bino--Higgsino scenario, $\Delta\Omega/\Omega\sim 15\%$
may be achieved~\cite{Baltz:2006fm}.
Whether similar precisions can be reached in the CPV-MSSM requires
careful investigation.

To emphasize the need to determine as completely as possible the
underlying parameters of the model, we stress again that we have
found examples where in the CPV-MSSM the relic density of the LSP
could be quite different as compared to that in the MSSM with
vanishing phases. Simply from a precise measurement of part of the
mass spectrum one could be lead to wrongly conclude that, for instance,
the model does not give a good dark matter candidate, or else
that some significant thermal production is necessary to explain
the observed number density. At the same time it is important to note 
that a computation of $\langle\sigma v\rangle$ at next-to-leading order 
will be necessary to achieve the required precision in the prediction of
$\Omega h^2$, see~\cite{Drees:1996pk,Bergstrom:1997fh,Boudjema:2005hb}. 
Last but not least we stress that in addition direct
or indirect detection of the CDM candidate will be indispensable
to pin down the dark matter in the Universe.

Finally we remark that we have not taken into account the
constraint arising from $b\rightarrow s\gamma$. The supersymmetric
corrections to the branching ratio for $b\rightarrow s\gamma$
depend mostly on the squark and gaugino/Higgsino sector as well as
on the charged Higgs. In the real MSSM, large corrections are
found at large values of $\tan\beta$, which we do not consider in
this paper. Large corrections might also arise in scenarios  with
a light charged Higgs. A detailed study of the impact of this
measurement in the CPV-MSSM is left for a future
work. \\ 

\noi
{\bf Note added:} While this paper was in preparation, the WMAP collaboration
published new results corresponding to three years of data taking.
The new WMAP+SDSS combined value for the relic density of dark matter is
$\Omega_{\rm CDM} h^2= 0.111^{+0.006}_{-0.011}$ at
$1\sigma$~\cite{Spergel:2006hy}.
This is only slightly below the value used in this paper, so our
conclusions do not change with the new data.

\section*{Acknowledgements}
\addcontentsline{toc}{section}{Acknowledgements}

We thank T.~Gajdosik and  W.~Porod for comparisons in the CPV-MSSM
and J.S.~Lee for his help with CPsuperH.
We also thank S.Y.~Choi for discussions.
This work was supported in part by GDRI-ACPP of CNRS
and by grants from the Russian Federal Agency for Science,
NS-1685.2003.2 and RFBR-04-02-17448.
S.K. is supported  by an APART (Austrian Programme for Advanced
Research and Technology) grant of the Austrian Academy of Sciences.
A.P. acknowledges the hospitality of CERN and LAPTH where some of the
work contained here was performed.

\input{appendix}


\end{document}

%% file: appendix.tex

\def\ma{m_A}
\def\mhf{m_{1/2}}
\def\m0{m_0}
\def\ra{\rightarrow}
\def\neuto {\tilde\chi_1^0}
\def\mneuto{m_{\tilde{\chi}_1^0}}
\def\stauo{\tilde\tau_1}
\def\staur{\tilde{\tau_R}}
\def\ser{\tilde{e_R}}
\def\smur{\tilde{\mu}_R}
\def\mt{m_t}
\def\mbmb{m_b(m_b)}
\def\mslr{m_{\tilde l_R}}

\def\nn              {\notag}
\def\bce             {\begin{center}}
\def\ece             {\end{center}}

\def\mbf             {\boldmath}
\def\noi             {\noindent}

\def\ti              {\tilde}

\def\a               {\alpha}
\def\b               {\beta}
\def\d               {\delta}
\def\D               {\Delta}
\def\g               {\gamma}
\def\G               {\Gamma}
\def\l               {\lambda}
\def\t               {\theta}
\def\s               {\sigma}
\def\S               {\Sigma}
\def\x               {\chi}

\def\sq              {\ti q}
\def\sqL             {\ti q_L^{}}
\def\sqR             {\ti q_R^{}}
\def\sf              {\ti f}

\def\st              {\ti t}
\def\sb              {\ti b}
\def\stau            {\ti \tau}
\def\snu             {\ti \nu}

\def\sqbar  {\Bar{\Tilde q}^{}}
\def\stbar  {{\Bar{\Tilde t}}}
\def\sbbar  {{\Bar{\Tilde b}}}

\def\PL              {P_L^{}}
\def\PR              {P_R^{}}

\def\Pp  {{\cal P}_{\!+}}
\def\Pm  {{\cal P}_{\!-}}

\def\ch              {\ti \x^\pm}
\def\chp             {\ti \x^+}
\def\chm             {\ti \x^-}
\def\nt              {\ti \x^0}

\newcommand{\msq}[1]   {m_{\ti q_{#1}}}
\newcommand{\msb}[1]   {m_{\ti b_{#1}}}
\newcommand{\mq}{\mbox{$m_{\tilde{q}}$}}
\newcommand{\mhp}      {m_{H^+}}
\newcommand{\msg}      {m_{\ti g}}

\def\sg              {{\ti g}}
\def\msg             {m_{\sg}}

\def\tW              {\t_W}
\def\tsq             {\t_{\sq}}
\def\tst             {\t_{\st}}
\def\tsb             {\t_{\sb}}
\def\tstau           {\t_{\stau}}
\def\tsf             {\t_{\sf}}
\def\sth             {\sin\t}
\def\cth             {\cos\t}
\def\cst             {\cos\t_{\st}}
\def\csb             {\cos\t_{\sb}}

\def\phmu            {\phi_\mu}
\def\phA             {\phi_A}
\def\phsf            {\varphi_{\!\sf}^{}}
\def\phst            {\varphi_{\ti t}^{}}
\def\phsb            {\varphi_{\ti b}^{}}
\def\phstau          {\varphi_{\ti\tau}^{}}

\def\onehf           {{\textstyle \frac{1}{2}}}
\def\oneth           {{\textstyle \frac{1}{3}}}
\def\twoth           {{\textstyle \frac{2}{3}}}

\def\rzw             {\sqrt{2}}

\def\BR              {{\rm BR}}
\def\mev             {{\rm MeV}}
\def\gev             {{\rm GeV}}
\def\tev             {{\rm TeV}}
\def\fb              {{\rm fb}}
\def\fbi             {{\rm fb}^{-1}}

\def\over            {\overline}
\def\MSbar           {{\overline{\rm MS}}}
\def\DRbar           {{\overline{\rm DR}}}
\def\DR              {{\rm\overline{DR}}}
\def\MS              {{\rm\overline{MS}}}



\newcommand{\smaf}[2] {{\textstyle \frac{#1}{#2} }}
\newcommand{\sfrac}[2] {{\textstyle \frac{#1}{#2}}}
\newcommand{\pif}      {\smaf{\pi}{2}}

\def\delr            {\!\stackrel{\leftrightarrow}{\partial^\mu}\!}

\begin{appendix}

\section{Interaction Lagrangian}

We here give the relevant sparticle interactions with other sparticles
and SM particles in the CPV-MSSM.
The Higgs boson interactions with sparticles and particles can be found 
in~\cite{Lee:2003nt}.

\subsection{Neutralino--neutralino--Z}

Neutralinos couple to $Z$ bosons via their Higgsino components:

\begin{equation}
   {\cal L}_{\nt\nt Z} = \frac{g}{4c_W}\,
   Z_\mu\, \overline{\nt_i}\,\gamma^\mu
   \left( O_{ij}^{''L}\,\PL + O_{ij}^{''R}\,\PR \right) \nt_j
\label{eq:ntntz}
\end{equation}
with $i,j=1,...,4$, $P_{L,R}^{}=\frac{1}{2}(1\mp\gamma^5)$ and
\begin{equation}
   O_{ij}^{''L} =
   (N_{i4}^{}N_{j4}^*-N_{i3}^{}N_{j3}^*) = -O_{ij}^{''R\,*} \,.
\end{equation}
One can also write Eq.~\eq{ntntz} as
\begin{equation}
   {\cal L}_{\nt\nt Z} = \frac{g}{4c_W} \,
   Z_\mu\, \overline{\nt_i}\,\gamma^\mu \left[\,
   i{\rm Im}(O_{ij}^{''L}) + {\rm Re}(O_{ij}^{''L})\,\gamma^5 \right] \nt_j \,.
\end{equation}

\subsection{\mbf Chargino--chargino--Z, $\gamma$}

The interaction of two charginos with electroweak gauge bosons is
\begin{equation}
   {\cal L}_{\ti\x^+\ti\x^+ Z} = \frac{g}{c_W}\,Z_\mu\,
   \overline{\ti\x^+_i}\,\gamma^\mu
   \left( O_{ij}^{'L}\,\PL + O_{ij}^{'R}\,\PR \right) \ti\x^+_j
   - e\,A_\mu\,\overline{\ti\x^+_i}\gamma^\mu\ti\x^+_i
\label{eq:chchz}
\end{equation}
with $i,j=1,2$,  and
\begin{eqnarray}
  O_{ij}^{'L} &=& -V_{i1}^{}V_{j1}^{*}
                  -\smaf{1}{2}V_{i2}^{}V_{j2}^{*}
                  +\delta_{ij}^{}\sin^2\tW\,,\\
  O_{ij}^{'R} &=& -U_{i1}^{*}U_{j1}^{}
                  -\smaf{1}{2}U_{i2}^{*}U_{j2}^{}
                  +\delta_{ij}^{}\sin^2\tW\,.
\end{eqnarray}

\subsection{Neutralino--chargino--W}

The neutralino--chargino--W interaction is described by
($i=1,2$; $j=1,...,4$)
\begin{equation}
   {\cal L}_{\ti\x\ti\x W} =
   g\; W^-_\mu \overline{\nt_j}\,\gamma^\mu
   \left( O_{ji}^{L}\,\PL + O_{ji}^{R}\,\PR \right) \ti\x^+_i
   + g\; W^+_\mu \overline{\ti\x^+_i}\,\gamma^\mu
   \left( O_{ji}^{L*}\,\PL + O_{ji}^{R*}\,\PR \right) \nt_j
\label{eq:ntchW}
\end{equation}
with
\begin{equation}
   O_{ji}^L = N_{j2}^{}V_{i1}^{*} - \smaf{1}{\sqrt{2}} N_{j4}^{}V_{i2}^{*}
   \quad {\rm and}\quad
   O_{ji}^R = N_{j2}^{*}U_{i1}^{} + \smaf{1}{\sqrt{2}} N_{j3}^{*}U_{i2}^{}\,.
\end{equation}

\subsection{Neutralino--fermion--sfermion}

The sfermion interaction with neutralinos is ($i=1,2$; $j=1,...,4$)
\begin{eqnarray}
  {\cal L}_{f\sf\nt}
  &=& g\,\bar f\,( f_{Lj}^{\sf}\PR + h_{Lj}^{\sf}\PL )\,\nt_j\,\sf_L^{} +
      g\,\bar f\,( h_{Rj}^{\sf}\PR + f_{Rj}^{\sf}\PL )\,\nt_j\,\sf_R^{}
      + {\rm h.c.}\nn\\
  &=& g\,\bar f\,( a^{\,\sf}_{ij}\PR + b^{\,\sf}_{ij}\PL )\,\nt_j\,\sf_i^{}
      + {\rm h.c.}
\label{eq:fsfnt}
\end{eqnarray}
where
\begin{eqnarray}
   a^{\,\sf}_{ij} &=& f_{Lj}^{\sf}\,R_{1i}^{\sf} +
                      h_{Rj}^{\sf}\,R_{2i}^{\sf},   \label{eq:aik}\\
   b^{\,\sf}_{ij} &=& h_{Lj}^{\sf}\,R_{1i}^{\sf} +
                      f_{Rj}^{\sf}\,R_{2i}^{\sf}.   \label{eq:bik}
\end{eqnarray}
The couplings $f_{L,R}^{\sf}$ and $h_{L,R}^{\sf}$ are
\begin{align}
  f_{Lj}^{\,\ti t} &= -\sfrac{1}{\sqrt 2}\,(N_{j2}
                        +\sfrac{1}{3}\tan\theta_W N_{j1}) \,, &
  f_{Lj}^{\,\ti b} &= \sfrac{1}{\sqrt 2}\,(N_{j2}
                        -\sfrac{1}{3}\tan\theta_W N_{j1}) \,, \\
  f_{Rj}^{\,\ti t} &= \sfrac{2\sqrt 2}{3}\,\tan\theta_W N_{j1}^*\,, &
  f_{Rj}^{\,\ti b} &= -\sfrac{\sqrt 2}{3}\,\tan\theta_W N_{j1}^*\,, \\
  h_{Rj}^{\ti t} &= -h_t^*\, N_{j4} = h_{Lj}^{\ti t*}\,, &
  h_{Rj}^{\ti b} &= -h_b^*\, N_{j3} = h_{Lj}^{\ti b*}\,
\end{align}
for stops and sbottoms, and
\begin{align}
  f_{Lj}^{\ti\tau} &= \sfrac{1}{\sqrt 2}\,(\tan\theta_W N_{j1}+N_{j2})\,,\\
  f_{Rj}^{\ti\tau} &= -\sqrt{2}\,\tan\theta_W N_{j1}^*\,,\\
  h_{Rj}^{\ti\tau} &= -h_\tau^*\, N_{j3} = h_{Lj}^{\ti\tau*}\,
\end{align}
for staus. In more general terms,
\begin{equation}
  f_{Lj}^{\,\ti f} = -\sqrt 2\left(
                     (e_f-I_{3L}^f)\tan\theta_W N_{j1} + I_{3L}^f N_{j2}
                     \right), \quad
  f_{Rj}^{\,\ti f} = \sqrt 2\,e_f\,\tan\theta_W N_{j1}^*\,.
\end{equation}

\subsection{Chargino--fermion--sfermion}

The sfermion interaction with charginos is ($i,j=1,2$)
\begin{eqnarray}
  {\cal L}_{f'\!\sf\ch}
  &=& g\,\bar u\,( -U_{j1}\PR + h_u\,V_{j2}^*\PL )\,\ti\x^+_j\,\ti d_L^{}
      + g\,\bar u\,( h_d^*U_{j2}\PR )\,\ti\x^+_j\,\ti d_R^{} + \nn\\
  & &
      g\,\bar d\,( -V_{j1}\PR + h_d\,U_{j2}^*\PL )\,\ti\x^{+c}_j\,\ti u_L^{}
      + g\,\bar d\,( h_u^*V_{j2}\PR )\,\ti\x^{+c}_j\,\ti u_R^{}
      + {\rm h.c.} \nn \\[1mm]
  &=& g\,\bar u\,( l_{ij}^{\,\ti d}\,\PR +
                   k_{ij}^{\,\ti d}\,\PL )\,\ti\x^+_j\,\ti d_i^{}
    + g\,\bar d\,( l_{ij}^{\,\ti u}\,\PR +
                   k_{ij}^{\,\ti u}\,\PL )\,\ti\x^{+c}_j\,\ti u_i^{}
      + {\rm h.c.}
\label{eq:fsfch}
\end{eqnarray}
where $u$ ($\ti u$) stands for up-type (s)quark and (s)neutrinos,
and $d$ ($\ti d$) stands for down-type (s)quark and charged (s)leptons.
The couplings $l$ and $k$ are
\begin{align}
  l_{ij}^{\ti t} &= -V_{j1} R_{1i}^{\ti t} + h_t^*\,V_{j2} R_{2i}^{\ti t}\,, &
  l_{ij}^{\ti b} &= -U_{j1} R_{1i}^{\ti b} + h_b^*\,U_{j2} R_{2i}^{\ti b}\,,
  \label{eq:elltb} \\
  k_{ij}^{\ti t} &= h_b\,U_{j2}^* R_{1i}^{\ti t} \,, &
  k_{ij}^{\ti b} &= h_t\,V_{j2}^* R_{1i}^{\ti b} \,,
  \label{eq:katb}
\end{align}
for stops and sbottoms and
\begin{align}
  l_{j}^{\ti\nu}  &= -V_{j1} \,, &
  l_{ij}^{\ti\tau} &= -U_{j1} R_{1i}^{\ti\tau} + h_\tau^*\,U_{j2}R_{2i}^{\ti\tau}\,, \\
  k_{j}^{\ti\nu}  &= h_\tau\,U_{j2}^* \,, &
  k_{ij}^{\ti\tau} &= 0 \,
\end{align}
for staus and sneutrinos.

\subsection{Sfermions with gauge bosons}

The sfermion interaction with photons is the same as in
the CP-conserving case:
\begin{equation}
  {\cal L}_{\ti f\ti f\g}
  = -iee_f\, A_\mu\,
    ( \sf_L^* \delr \sf_L^{} + \sf_R^* \delr \sf_R^{} )
  = -iee_f\,\d_{ij}\, A_\mu\: \sf_j^* \delr \sf_i^{} \,.
\end{equation}
The interaction with $Z$ bosons is given by
\begin{align}
  {\cal L}_{\ti f\ti fZ}
  & = -\frac{ig}{\cos\tW}\, Z_\mu\,
    (C_L^{}\,\sf_L^*\delr\sf_L^{} + C_R^{}\,\sf_R^*\delr\sf_R^{}) \nn\\
  & = -\frac{ig}{\cos\tW}
    \left( C_L^{}R_{1i}^{\sf}R_{1j}^{\sf\,*} +
           C_R^{}R_{2i}^{\sf}R_{2j}^{\sf\,*}\right)
    Z_\mu\,\sf_j^*\delr\sf_i^{}
\end{align}
with $C_{L,R}^{}=I_{3L,R}^f-e_f\sin^2\tW$. Note that there is only
a phase dependence for $i\not=j$.  \\
The interaction with $W$ bosons is given by
\begin{align}
  {\cal L}_{\ti f\ti f'W}
  &= -\frac{ig}{\sqrt{2}}\,
  (W^+_\mu\,\st_L^*\delr\sb_L^{} + W^-_\mu\,\sb_R^*\,\delr\st_R^{}) \nn\\
  &= -\frac{ig}{\sqrt{2}}\,
  (R_{1i}^{\ti b}R_{1j}^{\ti t\,*}\,W^+_\mu\,\st_j^*\delr\sb_i^{} +
   R_{1i}^{\ti t}R_{1j}^{\ti b\,*}\,W^-_\mu\,\sb_j^*\delr\st_i^{})
\end{align}
taking $\st\,\sb W$ as an example for simplicity.
The corresponding Feynman rules are obtained from
\begin{equation}
  A\delr B = A\,(\partial_\mu B) - (\partial_\mu A)\,B \quad \to \quad
  \sf_j^*\delr\sf_i^{} = i\,(k_i^{} + k_j^{})^\mu
\end{equation}
where $k_i^{}$ and $k_j^{}$ are the four--momenta of $\ti f_i^{}$
and $\ti f_j^{}$ in direction of the charge flow.

\end{appendix}